\documentclass[twocolumn]{aastex631}
\usepackage{natbib}
\usepackage{url}
\usepackage{amsmath}
\usepackage{appendix}
\usepackage{hyperref}
\usepackage{xcolor}
\usepackage{subfigure}
\usepackage{xspace}
\usepackage{soul}

\newcommand{\ha}{H\ensuremath{\alpha}\xspace}

\begin{document}%https://www.overleaf.com/project/621507092393f312c2bb2e2f

\title{Hubble Space Telescope Observations of Tadpole Galaxies Kiso 3867, SBS0, SBS1, and UM461 }

\correspondingauthor{Debra Meloy Elmegreen}
\email{elmegreen@vassar.edu}

\author[0000-0002-1392-3520]{Debra Meloy Elmegreen}
\affiliation{Vassar College, Dept. of Physics and Astronomy, Poughkeepsie, NY 12604}

\author[0000-0002-1723-6330]{Bruce G. Elmegreen}
\affiliation{IBM Research Division, T.J. Watson Research
Center, Yorktown Hts., NY 10598}

\author[0000-0001-8608-0408]{John S. Gallagher}
\affiliation{Dept. of Astronomy, Univ. of Wisconsin-Madison, Madison, WI 53706}

\author[0000-0002-4460-9892]{Ralf Kotulla}
\affiliation{Dept. of Astronomy, Univ. of Wisconsin-Madison, Madison, WI 53706},

\author{Jorge S\'anchez Almeida}
\affiliation{Instituto de Astrof\'isica de Canarias, C/
via L\'actea, s/n, 38205, La Laguna, Tenerife, Spain, and Departamento de Astrof\'isica,
Universidad de La Laguna}

\author{Casiana Mu\~noz-Tu\~n\'on}
\affiliation{Instituto de Astrof\'isica de Canarias, C/
via L\'actea, s/n, 38205, La Laguna, Tenerife, toSpain, and Departamento de Astrof\'isica,
Universidad de La Laguna}

\author[0000-0002-1653-5806]{Nicola Caon}
\affiliation{Instituto de Astrof\'isica de Canarias, C/
via L\'actea, s/n, 38205, La Laguna, Tenerife, Spain, and Departamento de Astrof\'isica,
Universidad de La Laguna}

\author[0000-0002-9946-4731]{Marc Rafelski}
\affiliation{Space Telescope Science Institute, 3700 San Martin Drive, Baltimore, MD 21218, USA}
\affiliation{Department of Physics and Astronomy, Johns Hopkins University, Baltimore, MD 21218, USA}

\author[0000-0003-3759-8707]{Ben Sunnquist}
\affiliation{Space Telescope Science Institute, 3700 San Martin Drive, Baltimore, MD 21218, USA}

\author[0000-0002-4917-7873]{Mitchell Revalski}
\affiliation{Space Telescope Science Institute, 3700 San Martin Drive, Baltimore, MD 21218, USA}

\author{Morten Andersen}
\affiliation{European Southern Observatory, Karl Schwarzschild Str. 2, 85748, Garching, Germany}

\begin{abstract}

Tadpole galaxies are metal-poor dwarfs with typically one dominant star-forming
region, giving them a head-tail structure when inclined. A  metallicity drop in
the head suggests that gas accretion with even lower metallicity stimulated the
star formation.  Here we present multiband HST WFC3 and ACS images of four
nearby ($<$ 25 Mpc) tadpoles, SBS0, SBS1, Kiso 3867, and UM461, selected for
their clear metallicity drops shown in previous spectroscopic studies.
Properties of the star complexes and compact clusters are measured.  Each galaxy
contains from 3 to 10 young stellar complexes with $10^3-10^5\;M_\odot$ of stars
$\sim3-10$ Myr old. Between the complexes, the disk has a typical age of $\sim3$
Gyr. Numerous star clusters cover the galaxies, both inside and outside the
complexes.  The combined cluster mass function, made by normalizing the masses
and counts before stacking, is a power law with a slope of $-1.12\pm0.14$ on a
log-log plot and the combined distribution function of cluster lifetime decays
with age as $t^{-0.65\pm0.24}$.  A comparison between the summed theoretical
Lyman continuum (LyC) emission from all the clusters, given their masses and
ages, is comparable to or exceeds the LyC needed to excite the observed
H$\alpha$ in some galaxies, suggesting LyC absorption by dust or undetected gas
in the halo, or perhaps galaxy escape.

\end{abstract} 

\section{Introduction}

Gas accretion onto galaxies drives their average star formation rates
\citep[SFRs;][]{finlator08,brooks09,combes14}, with excess local accretion
possibly connected to major bursts \citep{jorge14a,jorge14b,ceverino16}. Tadpoles
\citep{vdb,elmegreen05a}, or cometary galaxies \citep{mark,loose,cairos}, are
gas-rich dwarfs with one giant star-forming region that looks like the head of a
tadpole when the galaxy is viewed at some inclination; the rest of the disk
comprises the tail. This morphology is common at high redshift, where it is found
in  $\sim10$\% of resolved galaxies disk-l\citep{elmegreen05a,vdb,straughn,wind},
but it is rare locally at less than 0.2\% \citep{elm12}.  Based on simulations by
\cite{ceverino16} of low metallicity gas infall onto disk galaxies, which
typically triggers a burst of star formation near one end of the disk, we expect
that tadpole galaxies would appear disk-like if viewed face-on, with at least one
prominent star-forming region. Their baryon mass is dominated by gas (e.g.,
\cite{filho13}), and their stellar mass distribution tends to be triaxial with
axial ratios 1:0.7:0.4 \citep{putko}.

Tadpole galaxies in the local Universe (out to $\sim100$ Mpc) tend to be extremely
metal poor (XMP) \citep{morales}. Spectroscopic observations of 22 tadpoles
indicate their heads typically have a drop in metallicity by a factor of 3 to 10
compared with their tails, consistent with accretion of metal-poor gas
\citep{jorge13, jorge14a,jorge14b,jorge15,olmo17,lagos18}, a process that may be
generally important in XMP systems \citep[e.g.,][]{ekta10}. This relation between
excess SFR and drop in metallicity is observed in most low mass galaxies
\citep[e.g.,][]{2019ApJ...872..144H,2019ApJ...882....9S}.

One of the prototype tadpole galaxies is Kiso~5639. Its bright head is a large,
over-pressure, short-lived star-formation region with a rich population of star
clusters \citep{elm16} and a massive molecular cloud \citep{elmegreen18}. In our
previous HST imaging of it \citep{elm16} we  found the properties of the head to
be consistent with a model by \cite{ceverino16} where infalling gas forms a
low-metallicity starburst clump in the disk.

In this paper we present an HST imaging study in U, V, I, and H$\alpha$ of four
additional tadpole galaxies, SBS0 (KUG0940+544), SBS1 (SBS1129+576), Kiso~3867
(KUG0937+298) and UM461.  These were selected from previous tadpole spectroscopic
studies \citep{jorge15,jorge13} for showing clear metallicity drops in the
star-forming heads, as discussed further below.  Galaxy coordinates, distances,
masses, and rotational speeds are listed in Table \ref{tab1}.

Our objective is to study the properties of the dominant XMP star forming
complexes and how they correlate with the characteristics of host galaxies.  We
are also interested in the populations of young star clusters formed in these
bursts, and in their mass and age distributions. For example, the localized
starbursts in these tadpoles could have been triggered by infalling gas streams,
and in other galaxies where this has happened, such as NGC 1569 \citep{johnson12}
and NGC 5253 \citep{turner15}, there are superstar clusters near the points of
impact. We would like to know if there are outlier superstar clusters in these
tadpoles or only a smooth extrapolation of the usual power-law cluster mass
function \citep{elmegreen97}.

\begin{deluxetable}{cccccc}
%[h!]
\tabletypesize{\scriptsize}\tablecolumns{6} \tablewidth{0pt} \tablecaption{Observed Galaxies}
\tablehead{\colhead{Galaxy} &\colhead{RA} &\colhead{Dec} &\colhead {D$_{GSR}$\tablenotemark{a}} &\colhead{log Mass\tablenotemark{b}} & \colhead{V$_{rot}$sin i\tablenotemark{b}}  \\
\colhead{} &\colhead{(J2000)} &\colhead{(J2000)} &\colhead{(Mpc)} &\colhead{(M$_{\odot}$)} &\colhead{(km s$^{-1}$)}}
\startdata
SBS0      & 9:44:17  &  +54:11:29  &  24.7  &  $7.1 \pm 0.05$  & $15  \pm 1.2$  \\
SBS1      & 11:32:02 &  +57:22:45  &  24.2  &  $7.5 \pm 0.5$   & $36  \pm 4$    \\
%Kiso 3867 & 9:40:13  &  +29:35:30  &   6.7  &  $7.2 \pm 0.04$  & $20  \pm 10$   \\
Kiso 3867 & 9:40:13  &  +29:35:30  &   6.7  &  $6.3 \pm 0.2$  & $20  \pm 10$   \\
UM461     & 11:51:33 &  -02:22:22  &  13.6  &  $6.7 \pm 1.0$   & $8.6 \pm 0.7$
\enddata
\label{tab1}
\tablenotetext{a}{Galactocentric distances are from the NASA Extragalactic Database, ned.ipac.caltech.edu}
\tablenotetext{b}{Stellar masses and velocities for SBS0, SBS1, and UM461 from \cite{jorge15}, and
stellar mass for Kiso 3867 from \citet{filho13}
and velocity from \cite{jorge13}}
\end{deluxetable}

All four galaxies have lower metallicities in the bright star-forming regions than elsewhere. The metallicity drops in SBS0 (J094416.6+541134.4), SBS1 (J113202.4+572245.2), and UM461 (J115133.3-022221.9) were shown by \cite{jorge15} using the long-slit spectrograph OSIRIS at the 10.4 m Gran Telescopio Canarias\footnote{http://www.gtc.iac.es/GTChome.php}.
The metallicity drop in Kiso 3867 was shown by \cite{jorge13} using the spectrograph IDS of the 2.5 m Isaac Newton Telescope at the Observatorio del Roque de Los
Muchachos.

SBS0, SBS1 and Kiso 3867 are morphological tadpoles, which suggests they are inclined disks.  UM461 is presumably more face-on \citep[inclination $\sim$30$\pm$ 10$^{\circ}$;][]{vanzee98}. UM461 is also unusual in having 2 luminous, compact, young stellar regions. The H{\sc i} line map of UM461 by \cite{vanzee98} shows an extended H{\sc i} disk, typical of blue compact dwarf (BCD) systems, with no indication of a tidal interaction with the nearby UM462 system. The outer H{\sc i} shows evidence for filamentary structures. Similar features in other BCDs are suggested as possible indications of ongoing gas accretion (e.g., IC10; \cite{wilcots98, namumba19}, but see also comments by \cite{lelli14}).

\cite{ekta06} mapped SBS1 in the H{\sc i} 21~cm line using the Giant Metrewave Radio Telescope (GMRT). SBS1 is interacting with its more massive companion galaxy SBS~1129+577 and an H{\sc i} bridge joins the two galaxies. The interaction, however, seems unlikely to be the source of low metallicity material in SBS1 since the more massive and optically brighter SBS~1129+577 galaxy is expected to have higher metallicity than SBS1. Kiso~3868 (AGC~194068) was observed by \cite{pustilnik07} using the Nançay telescope with Arecibo in the ALFALFA H{\sc i} survey \citep{haynes18} with
log(M$_{H{\sc I}}$/M$_{\odot})=7.6$ and M$_{H{\sc I}}$/L$_B \approx$3 \citep{durbala20}. The SBS0 system has only a weak upper limit on the H{\sc i} mass that still allows it to be a gas-rich system \citep{filho13}. In summary, the tadpoles in our sample resemble typical low luminosity BCDs in terms of their H{\sc i} properties and include the SBS1 interacting system and 3 examples of apparently single dwarf tadpole galaxies.

Our observations and data reduction methods are presented in Section
\ref{sect:reduce}, the analysis of large star-forming clumps, which include clusters and other young stars, is discussed in Section \ref{sect:clump}, the star clusters themselves are considered in Section \ref{sect:cluster}, cluster mass and age distribution functions are in Section \ref{massfunction}, H$\alpha$ emission is in Section \ref{sect:Ha}, and the conclusions are in Section \ref{sect:conc}.

\section{Observations}\label{sect:reduce}
The four tadpoles in our sample were imaged with the Hubble Space Telescope for Cycle 27 GO proposal 15860 using the WFC3/UVIS and ACS cameras between November 2019 and April 2021.
The filters used were F390W (U), F547M (V), and F657N (H$\alpha$) with WFC3/UVIS, and F814W (I) with ACS, with 1 orbit for each filter per galaxy, for a total of 16 orbits.

The 5$\sigma$ detection limits in AB mag from the HST exposure time calculator (ETC) for point sources are 27.2, 26.6, and 27.1 for filters F390W, F547M, and F814W, respectively. The 5$\sigma$ limit for the H$\alpha$ filter F657N is $4.5 \times 10^{-17}$~erg~cm$^2$~s$^{-1}$.
%\jorge{\st{Does $\sigma$ refers to pixel-to-pixel variation? Say it explicitly}}

The WFC3/UVIS images were custom calibrated using the routines developed in \citet{Rafelski:2015} and \citet{Prichard:2022}. First, the images were corrected with a pixel based correction for charge transfer efficiency (CTE) degradation to account for the radiation damage over time \citep{MacKenty:2012}. We utilized the new improved CTE algorithm \citep{Anderson:2021} with reduced noise amplification and improved corrections. We used the new co-temporal UVIS darks to apply a custom hot pixel mask using a variable threshold as a function of distance from the amplifier readout to detect a constant number of hot pixels across the CCDs\footnote{\url{https://github.com/lprichard/HST_FLC_corrections}}. We flagged readout cosmic rays, which are cosmic rays that land on the CCD while being read out, and are otherwise over-corrected by the CTE code which would result in negative divots in the images. We then equalized all four amplifier regions to produce images with constant background levels\footnote{\url{https://github.com/bsunnquist/uvis-skydarks}}. Finally, due to the small number of exposures per filter, we applied a custom cosmic ray rejection based on the \href{https://www.astropy.org}{Astropy} implementation of the \textsc{LACOSMIC} procedure\footnote{\url{https://github.com/mrevalski/hst_wfc3_lacosmic}} \citep{vanDokkum:2001, McCully:2019}.

The resulting custom calibrated science exposures were aligned to GAIA DR2 \citep{Gaia:2018} with \textsc{TweakReg} and combined with \textsc{AstroDrizzle} with a 0.8 pixel fraction at a final pixel scale of 0.03$''$. Root mean square (RMS) uncertainty maps were created from the inverse variance (IVM) images as 1/sqrt(IVM), and were corrected for correlated pixel noise \citep{Fruchter:2002}.

%fig1
\begin{figure*}
%\center{\includegraphics[scale=0.83]{f1.pdf}}
\center{\includegraphics[scale=0.6]{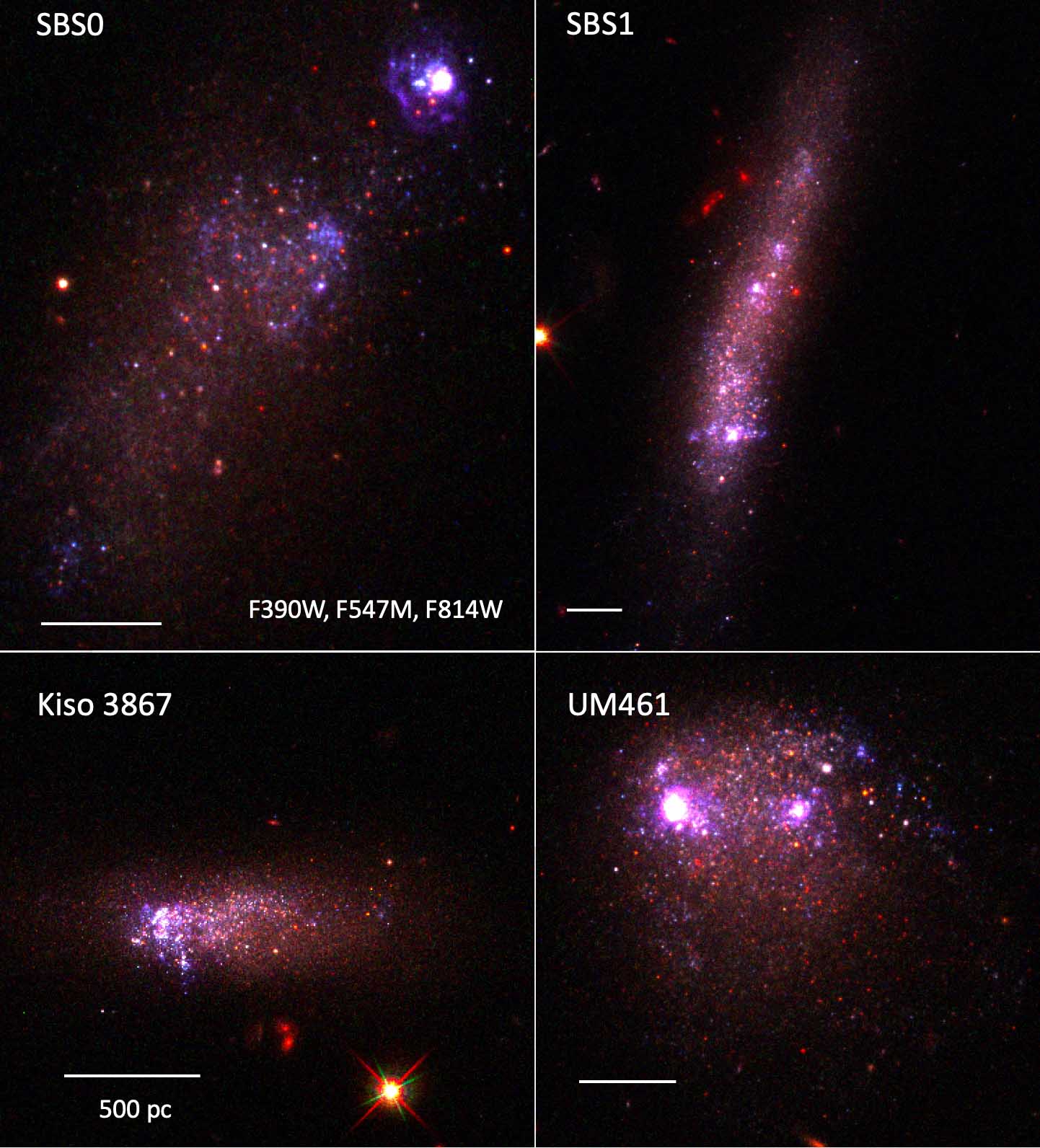}}
%\center{\includegraphics[width=\textwidth]{f1_cut.jpg}}
\caption{Tadpole images from HST WFC3 and ACS observations, using filters F390W, F574M, and F814W for blue, green, red. The galaxies are labeled. The line indicates 500 pc; the HST images have pixel scales of $0.03^{\prime\prime}$.  N is up, E to the left. }
\label{color}
\end{figure*}

%fig2
\begin{figure*}
%\center{\includegraphics[scale=.83]{f2.pdf}}
%\center{\includegraphics[width=\textwidth]{4gals__logscale.pdf}}
\center{\includegraphics[width=\textwidth]{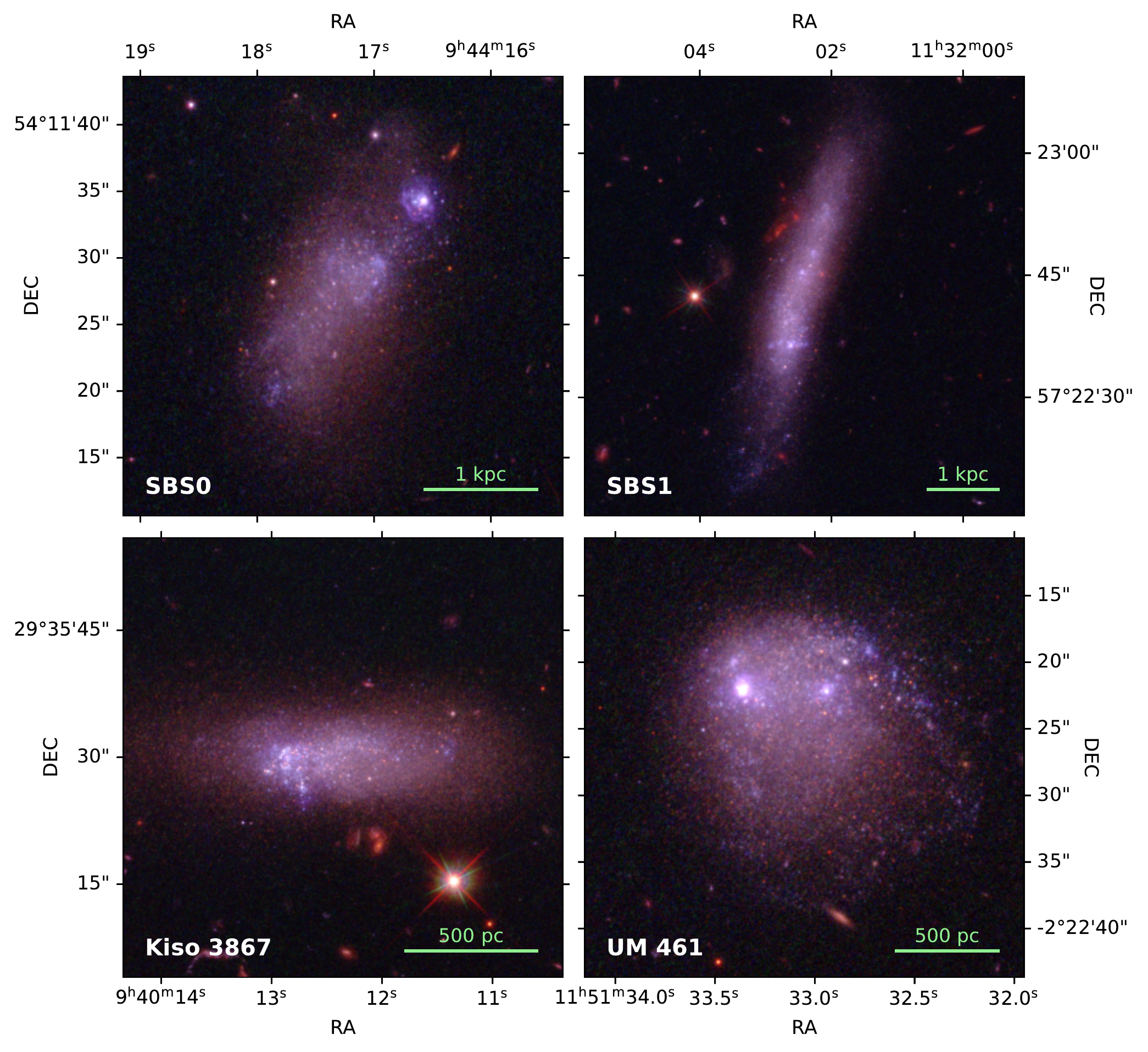}}
%\caption{Alternative to Figure 2 -- also using log-scale, %but with coordinate axes as suggested in the comments. %\newline
\caption{Color composite figures as in Figure \ref{color} using filters F390W, F574M, and F814W for blue, green, red, but here with logarithmic stretches. N is up, E to the left.}
\label{colorlog}\end{figure*}

Color composite figures were made in the Image Reduction and Analysis Facility (IRAF)
using RGB in DS9, with the F390W (for blue), F547M (for green), and F814W (for red) images. These are shown in
Figures \ref{color} and \ref{colorlog} for linear and logarithmic stretches, respectively. As traced by the relative blue colors, star formation is prominent in the heads of SBS0 and Kiso 3867.  The head is not as prominent in SBS1, which also has some bright star-forming regions along its tail. UM461 has two large star-forming regions in the E and W portions in the northern half of the galaxy. Star clusters are evident in several regions throughout the galaxies, discussed below. The logarithmic stretches %\bruce{to Jorge: like mag images but without the -2.5 product and without the zero point etc, so not really mags. But log stretches for imagery is common and I don't think it needs an additional comment.}
show extended emission well beyond the brightest parts of the disks. In UM461, the southwest part of the galaxy is extended away from the star-forming peaks and a ridge containing young stellar populations extends westwards from the north side of the system.

To quantify the continuum associated with the H$\alpha$ images, a linear interpolation of the F547M and F814W filters was performed.
First, each broad-band image was normalized to the H$\alpha$ image by dividing it by the ratio of the corresponding
PHOTFLAM values stored in the image header. This bring the three images into a common photometric system,
where a source with constant flux per unit wavelength would have exactly the same number of counts.
Aperture photometry was then carried out for those stellar objects in the field visible in the three
filters, excluding those objects too faint in the H$\alpha$ image to give meaningful results, and
those that appeared to be saturated in the F814W frame. There were only two or three useful stellar
objects for each galaxy, so that a continuum fit on a galaxy by galaxy base was not feasible.
Instead, the results for the four galaxies were merged together, for a total of 10 useful sources,
and the best fit was computed to the relation:
\begin{equation}
f(F657N) = a\times f(F814W) + b\times f(F547M) % *--> \times Jorge
\label{Eq:contHa}
\end{equation}
where $f$ are the fluxes (on the normalized images) derived from the aperture photometry.
The best-fit coefficients are: $a = 0.524 \pm 0.139$ and $b = 0.456 \pm 0.130$, with the
uncertainties computed using the bootstrap method.
Equation~(\ref{Eq:contHa}) assumes the calibration stars to have negligible H$\alpha$ emission and an average spectral energy distribution (SED) similar to the stellar parts of the galaxies.

\begin{figure*}
\center{
\includegraphics[width=1.00\textwidth]{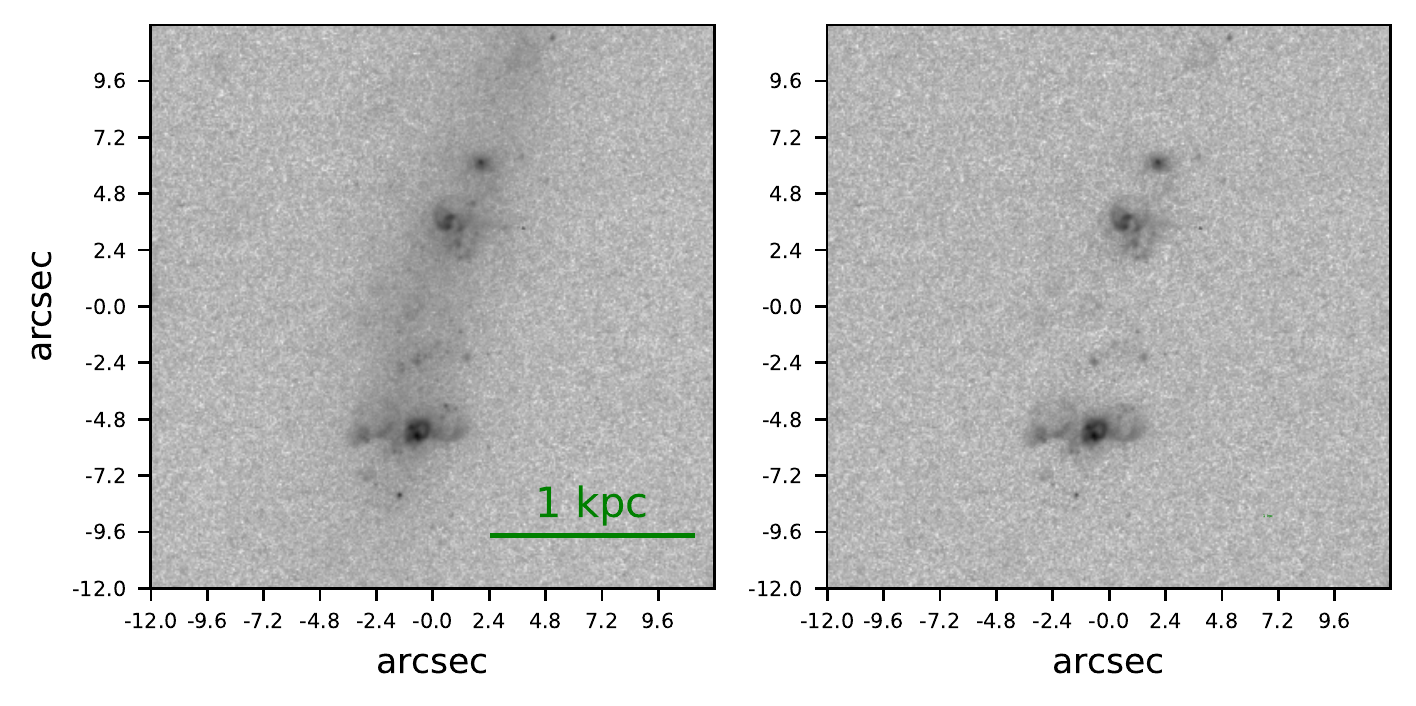}
}
\caption{H$\alpha$ image of SBS1 before (left) and after (right) continuum subtraction, as discussed in Section 3.}
\label{SBS1_Ha_contsub}\end{figure*}

As an example, Figure~{\ref{SBS1_Ha_contsub} shows the H$\alpha$ image of SBS1 before and after continuum subtraction.  The H$\alpha$ image was calibrated considering the throughput of the whole optical system, including the F657N filter, at the observed H$\alpha$ wavelength.}

%fig3

%fig4
\begin{figure*}
%\center{\includegraphics[scale=0.65]{f3.pdf}}
\center{\includegraphics[scale=0.3]{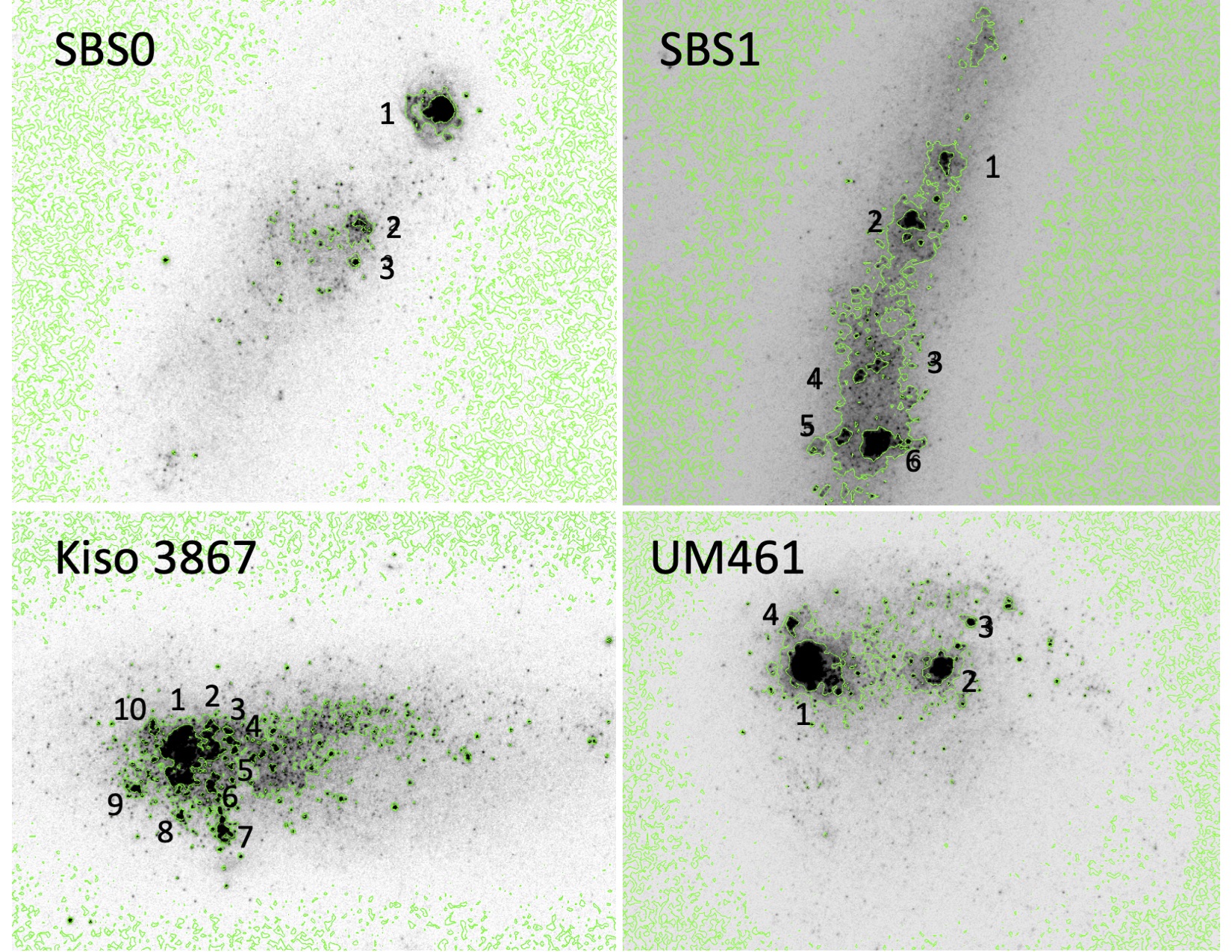}}
\caption{F390 images are shown for each galaxy, with green isocontours corresponding to 20$\sigma_{sky}$  in the galaxies and $\sigma_{sky}$ in the outer parts. The  flux for each  star-forming clump, marked in the figure, is integrated within the contour. Measurements were made in boxes drawn around these contours for clumps larger than 20 pixels diameter, as labeled. }\label{contour}\end{figure*}

%fig5
\begin{figure*}
%\center{\includegraphics[scale=0.55]{f4.pdf}}
\includegraphics[scale=0.2]{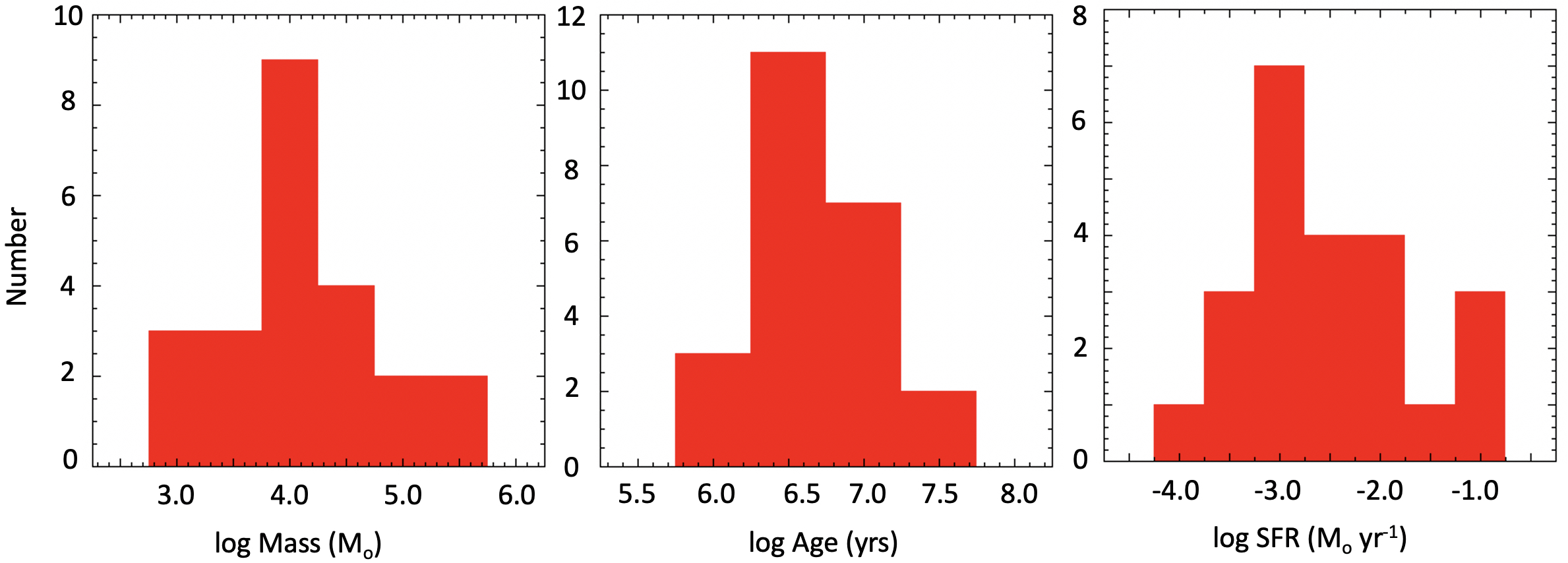}
%\center{\includegraphics[scale=0.15]{f5.jpg}}
\caption{Histograms are shown for the ages, masses, and star formation rates (based on age/mass) for the measured clumps, indicated in Figure \ref{contour}, based on the values provided in Table 2. }\label{histclump}\end{figure*}

\begin{deluxetable}{lccccc}
\tabletypesize{\scriptsize}\tablecolumns{6} \tablewidth{0pt} \tablecaption{Clump Properties}
\tablehead{
\colhead{Clump}&
\colhead{log Mass\tablenotemark{a}} &
\colhead{log Age} &
\colhead{A$_V$}&
\colhead{log SFR}&
\colhead{Surf Dens}
\\
\colhead{}&
\colhead{M$_{\odot}$} &
\colhead{yr} &
\colhead{mag}&
\colhead{M$_{\odot}$ yr$^{-1}$}&
\colhead{M$_{\odot}$ pc$^{-2}$}
}
\startdata
SBS0   	&					&					&					&		               \\
1	        &	$5.3 \pm 0.24$	&	$6.4 \pm 0.31$	&	$0.48 \pm 0.31$ &	$-1.0$&$11.3$	\\
2	        &	$4.2 \pm 0.21$	&	$6.3 \pm 0.24$	&	$0.54 \pm 0.24$	&	$-2.1$&$3.9$	\\
3	        &	$3.8 \pm 0.30$	&	$6.5 \pm 0.42$	&	$0.49 \pm 0.40$	&	$-2.7$&$4.7$	\\
SBS1	    &					&					&					&                   \\
1  	        &	$4.2 \pm 0.25$	&	$6.3 \pm 0.33$	&	$0.47 \pm 0.34$	&	$-2.1$&$2.5$	\\
2	        &	$4.6 \pm 0.29$	&	$6.5 \pm 0.47$	&	$0.52 \pm 0.39$	&	$-2.0$&$3.3$	\\
3	        &	$4.3 \pm 0.32$	&	$6.9 \pm 0.68$	&	$0.77 \pm 0.53$	&	$-2.5$&$2.9$	\\
4	        &	$4.2 \pm 0.29$	&	$7.1 \pm 0.68$	&	$0.75 \pm 0.56$	&	$-2.8$&$4.3$	\\
5	        &	$3.8 \pm 0.01$	&	$6.1 \pm 0.13$	&	$0.05 \pm 0.05$	&	$-2.3$&$1.1$	\\
6	        &	$5.0 \pm 0.11$	&	$6.2 \pm 0.20$	&	$0.31 \pm 0.22$	&	$-1.2$&$3.3$	\\
Kiso 3867	&				    &  	                &	                &                   \\
1 	        &	$4.7 \pm 0.32$	&	$6.8 \pm 0.56$	&	$0.66 \pm 0.47$	&	$-2.0$&$6.7$	\\
2        	& 	$4.2 \pm 0.31$	&	$7.0 \pm 0.69$	&	$0.78 \pm 0.54$	&	$-2.7$&$6.1$	\\
3	        &	$3.6 \pm 0.26$	&	$7.4 \pm 0.69$	&	$1.13 \pm 0.62$	&	$-3.8$&$13.8$	\\
4        	&	$3.4 \pm 0.34$	&	$6.7 \pm 0.51$	&	$0.59 \pm 0.47$	&	$-3.3$&$4.5$	\\
5	        &	$3.2 \pm 0.32$	&	$6.5 \pm 0.42$	&	$0.58 \pm 0.40$	&	$-3.3$&$6.3$	\\
6	        &	$4.0 \pm 0.29$	&	$7.2 \pm 0.80$	&	$0.97 \pm 0.65$	&	$-3.2$&$10.0$	\\
7        	&	$3.9 \pm 0.33$	&	$6.6 \pm 0.45$	&	$0.60 \pm 0.42$	&	$-2.7$&$4.6$	\\
8	        &	$3.1 \pm 0.10$	&	$6.2 \pm 0.19$	&	$0.27 \pm 0.19$	&	$-3.1$&$4.0$	\\
9	        &	$3.7 \pm 0.30$	&	$7.0 \pm 0.72$	&	$0.82 \pm 0.59$	&	$-3.3$&$13.5$	\\
10	        &	$3.2 \pm 0.11$	&	$6.3 \pm 0.20$	&	$0.28 \pm 0.15$	&	$-3.1$&$2.7$	\\
UM461	    &					&			        & 	                & 		            \\
1	        &	$5.6 \pm 0.30$	&	$6.7 \pm 0.56$	&	$0.57 \pm 0.43$	&	$-1.1$&$14.7$	\\
2        	&	$4.8 \pm 0.29$	&	$6.5 \pm 0.42$	&	$0.57 \pm 0.36$	&	$-1.7$&$6.8$	\\
3	        &	$4.4 \pm 0.27$	&	$7.4 \pm 1.07$	&	$0.99 \pm 0.76$	&	$-3.0$&$26.5$	\\
4	        &	$4.0 \pm 0.32$	&	$6.8 \pm 0.59$	&	$0.64 \pm 0.49$	&	$-2.8$&$4.2$	\label{tab2}
\enddata
\tablenotetext{a}{The masses and ages of the clump regions corresponding to
rectangles as described in the text, measured at contours 20$\sigma_{sky}$. The star formation rate is assumed to be constant over a time span given by the age.}
\end{deluxetable}

\section{Large-scale Star-forming Clumps}\label{sect:clump}

Photometric measurements were made of the large-scale star-forming complexes, or clumps,
larger than 20 pixels ($0.6^{\prime\prime}$)
in diameter for each galaxy using the task
{\it imstat} in IRAF. Boxes were defined around the clumps based on isophotal contours
in the F390W image at $20\sigma_{sky}$,
 labeled in Figure \ref{contour}. The background was subtracted
from each clump measurement to derive the clump magnitudes. Photometric zeropoints
calculated from the WFC3
website\footnote{\url{http://www.stsci.edu/hst/wfc3/phot_zp_lbn}} and ACS
website\footnote{https://www.stsci.edu/hst/instrumentation/acs/data-analysis/zeropoints}
were used to convert counts to AB mag.

Ages and extinctions for the tadpole clumps were determined by fitting stellar population models
to the observed spectral energy distributions, using the (U-V) and (V-I) colors. The observed colors were
corrected for background by subtracting in each
filter the counts for nearby fields from the counts in the clumps. Uncertainties in the
corrected counts were set equal to the square roots of the sums of the standard
deviations in the clump and background counts, weighted by the number of pixels for
each, and the relative uncertainties were taken to be these corrected uncertainties
divided by the total background-subtracted counts in the clump. The relative uncertainties
in the colors were taken to be the square roots of the sum of the squares of the
relative uncertainties in each passband. On average, the uncertainties in the colors,
converted to magnitudes, were 0.69 mag for (U-V) and 0.71 mag for (V-I).

The intensities were modelled with single stellar population models at 0.2 solar abundance and a Kroupa IMF in \cite{bruzual}
using the full model spectrum at each trial age multiplied by $\exp(-\tau)$ for
wavelength-dependent dust opacity $\tau$ and by the wavelength-dependent filter
throughput. We used extinction curves from \cite{calz00} and \cite{leith02} with
$A_V$ ranging from 0 to 4 mags in steps of 0.1 mag. Throughputs for the UVIS filters
are from the WFC3
website\footnote{\url{http://www.stsci.edu/hst/wfc3/ins_performance/throughputs/Throughput_Tables}}.
The two observables, (U-V) and (V-I), were thus converted to two physical quantities, age and extinction.

The resulting intensities for each passband were integrated over time for population ages back to some start time, which is identified with the age of the region. This integration assumes a constant star formation
rate. The population models assume a Chabrier IMF with a metallicity of 0.2 solar. Trial models varied
the population age from $10^6$ to $10^{10}$ years in 16 equal steps of $\log({\rm age})$. For
each age and extinction, the model colors were set equal to $-2.5$ times the log of the ratio of intensities.

The best simultaneous fit for age and extinction was determined by minimizing the
differences between the model and observed colors. One method weighted all of the
age and extinction values by $\exp(-\chi^2/2)$ within a limited range of $\chi$ for
$\chi^2$ equal to the sum of the squares of the differences between the observed and model colors divided
by their respective uncertainties. We then used the average of these weighted
values as the fitted solution. A second method took the $\chi^2$ weighted average
of the age plus extinction values with the lowest root mean square differences between
the model and observed colors, binning these differences in steps of 0.1 mag.
The uncertainties for both methods were taken to be the rms uncertainties in
these averages. Both methods gave about the same results, so here we report those
from the latter method.

After the ages and extinctions were fit from the colors, the masses of the clumps
were determined by setting the observed absolute intensity in the I band to the
model absolute intensity, using the mass versus age tabulation in \cite{bruzual}.

Table \ref{tab2} lists the masses, ages, and extinctions derived from model fits to the photometry for the
large-scale clumps in the galaxies, as well as SFR and stellar surface density. The star formation rates listed are the ratios of the
 total stellar masses formed (not the present masses) divided by the ages.

 The clump average log (Mass) ($M_{\odot}$) is 4.1$\pm 0.7$.  The total log mass in clumps in each galaxy is 5.4, 5.3, 5.0, and 5.7, in listed galaxy order. The clump average log (Age) (yrs) is 6.7$\pm 0.4$, with a range from 6.1 to 7.4.
Extinction A$_V$ ranges from 0.3 to 1.1 mag  for
all of the clumps, with an average of 0.6$\pm 0.2$ mag. The log SFR average for the clumps
is -2.5$\pm0.8$.
Histograms showing the distributions for all 4 galaxies for mass, age, and SFR are
shown in Figure \ref{histclump}. The sizes of the clumps (from the square root of the area of the clumps) range from 16 to 162 pc, averaging 65$\pm 47$ pc. Stellar surface densities range from 1.1 to 26.5 $M_{\odot}$ pc$^{-2}$, with an average of 7.0$\pm 5.7$ $M_{\odot}$ pc$^{-2}$. For comparison, the background surface densities range from 22 to 97 $M_{\odot}$ pc$^{-2}$, with an average of 43.5$\pm 29.7$ $M_{\odot}$ pc$^{-2}$.

For comparison, \cite{lagos11} used Gemini/NIRI J, H, K images of UM461 to measure star complexes and modeled their masses and ages. Their clumps 2 and 5 in their figure 4 correspond to our biggest clumps 1 and 2, respectively. Their (log (Age), log (Mass)) for these are (6.3, 5.6) and (6.4, 4.6) compared with our  values of (6.7, 5.6) and (6.5, 4.8), so the infrared and optical results are similar. Their clumps 1 and 9 are similar to our 4 and 3 also. In addition, they find several clumps with log mass ($M_{\odot}$) $\sim 4.6$ that do not stand out in our optical images. In particular, their clump 4 has no obvious optical counterpart, indicating obscuration south of our clump 1.

Our simultaneous fitting of age and
extinction to only two broadband colors, (U-V) and (V-I), produces uncertainties in both quantities, especially since they compensate for each other with older ages giving redder colors like higher extinctions. To check the inaccuracies, we forced all the fits for star complexes to a constant extinction of $A_V=0.1$ mag, which is less than the fitted values. As expected, the ages increase and the masses decrease a little. For the four galaxies SBS0, SBS1, Kiso 3867 and UM461, respectively, the averages in the log (Mass) ($M_\odot$) decrease from 4.4 to 4.1, 4.4 to 4.2, 3.7 to 3.5, and 4.7 to 4.4. The averages in the log (Age) (yr) increase from 6.4 to 6.8, 6.5 to 6.9, 6.8 to 7.4, and 6.9 to 7.6. These represent a 5\% decrease in log (Mass) and a 8\% increase in log (Age). The star formation rates obtained by the ratios of mass to age decrease with these changes, by an average factor of 5.8 (-0.76 in the log) for all galaxies if $A_V$ is forced to equal 0.1 mag.

\begin{figure}
% \center{\includegraphics[scale=0.35]{f5.pdf}}
 \center{\includegraphics[scale=0.4]{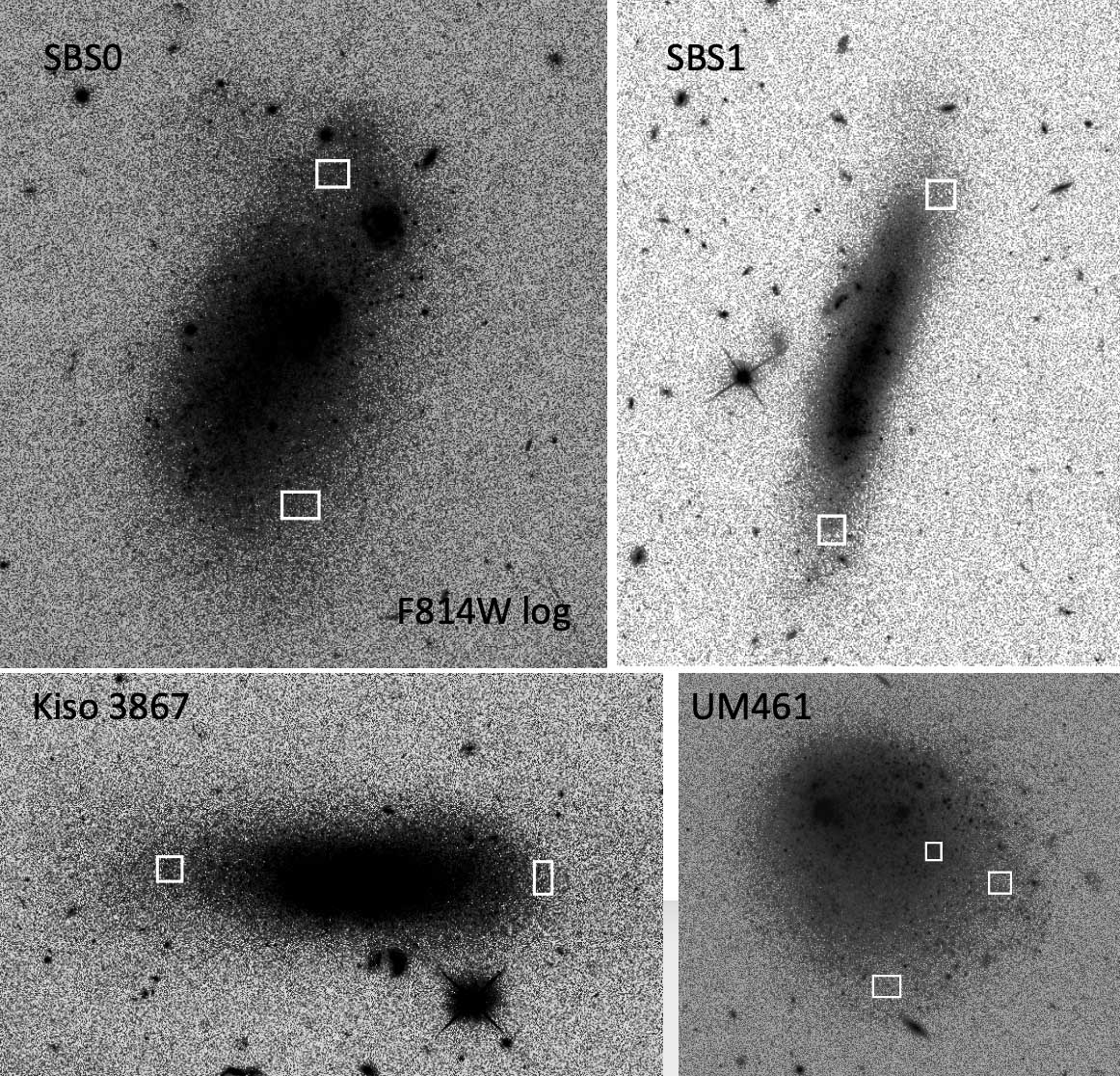}}
\caption{Logarithmic stretches of the F814W images for each galaxy are shown. N is up, E is to the left. The white boxes correspond to the non-clump regions measured in each galaxy.}\label{inter}\end{figure}

\begin{figure}
%\center{\includegraphics[scale=0.35]{f6.pdf}}
\center{\includegraphics[scale=0.15]{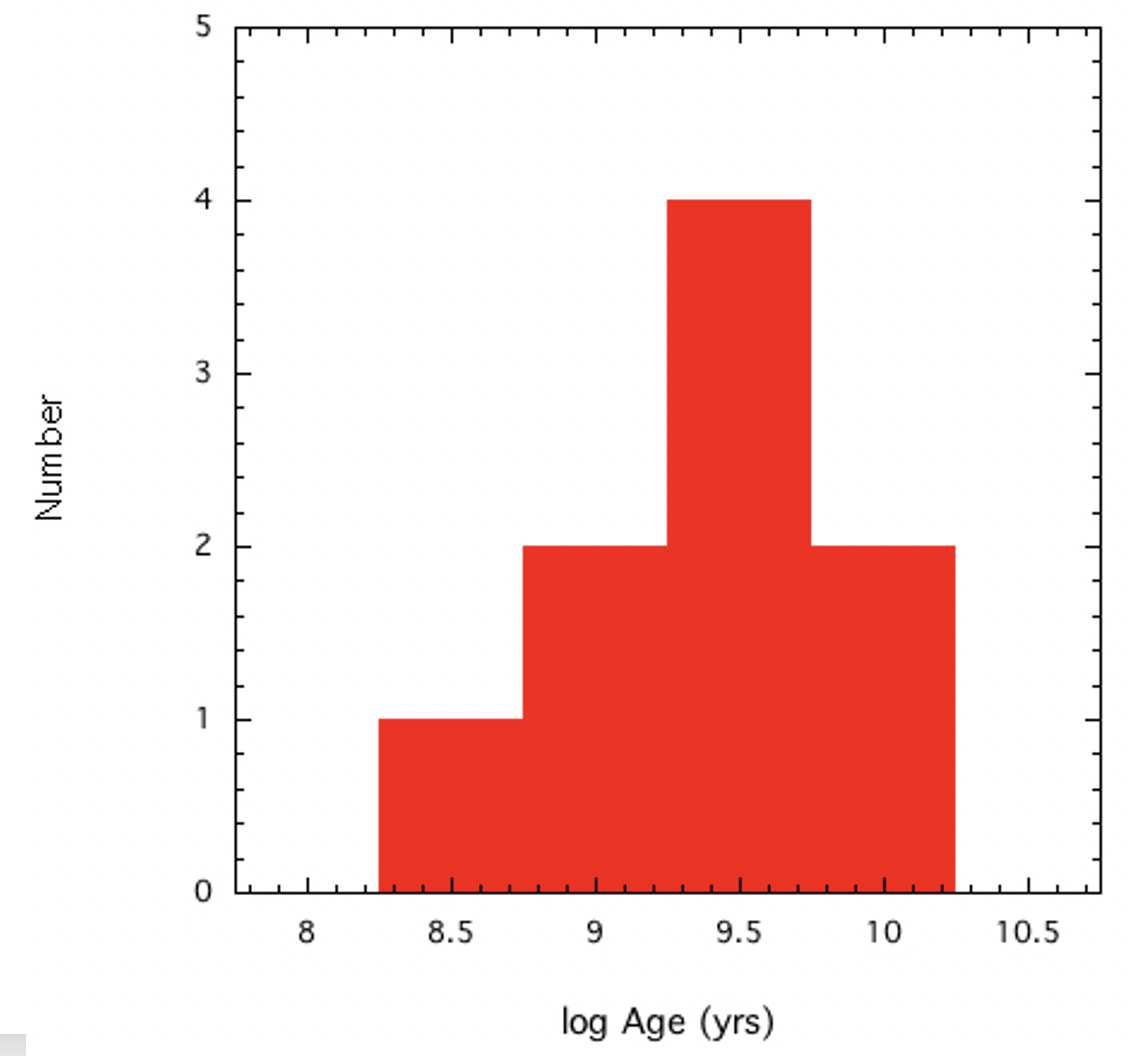}}
\caption{A histogram is shown for the ages of the interclump regions measured in the outer parts of the galaxies, as indicated by rectangles in Figure \ref{inter}. }\label{interage}\end{figure}

The extended diffuse components of the galaxies are expected to be older than the clumps, which photometric fits confirm. These outer diffuse regions are highlighted in the logarithmic stretch images in F814W shown in Figure \ref{inter}. SBS0 shows extensions in the NE and SW, SBS1 shows extensions in the N and S, Kiso 3867 shows extensions in the W and particularly in the E, and UM461 shows a long extension in the SW. Rectangular boxes devoid of obvious clumps or star clusters were measured for background subtraction of the clumps, described earlier, and to determine surface brightnesses and ages. A histogram of their ages is shown in Figure \ref{interage}; their average log (Age) (yrs) is 9.3$\pm 0.4$.

%fig6

%fig7

\section{Star Clusters}\label{sect:cluster}

\subsection{Cluster measurements}
To study star clusters in each galaxy, we extracted photometry for all compact, point source-like sources using SourceExtractor \citep{Bertin}. We varied the background matching and subtraction parameters across a wide range of parameters, from finely-grained to large-scale to account for systematic uncertainties in estimation the underlying galaxy contribution. Final photometry was obtained in circular apertures, detected in the F814W band, and fixed in position across all different bandpasses to extract physically consistent colors and thus spectral energy distributions across all bands. Photometry variations due to the different background modeling strategies were accounted for alongside the photometric uncertainties as total uncertainties. To extract physical parameters for each cluster we used $\chi^2$ minimization to fit each observed SED to a grid of GALEV stellar population models of fixed metallity of $\mathrm{[Fe/H]}=-0.7$ and a Kroupa-IMF with an upper mass limit of 100 M$_\odot$, with stellar population age, dust reddening using the Calzetti extinction law, and stellar mass as free parameters to be optimized. While we included data from the F659N band in the photometry, we chose to not include data in this band for the SED fitting due to the more extended nature of the H$\alpha$ emission.

Note that due to the forced photometry with detection in the F814W band, we may obtain no significant detection in the shorter wavelength bands for older and/or low-mass clusters. We therefore require a valid $5\sigma$ detection (combined uncertainty $<0.2$ mag) in the F390W band for a cluster to be include in the subsequent analysis.

\begin{figure*}
    \centering
    \includegraphics[width=\textwidth]{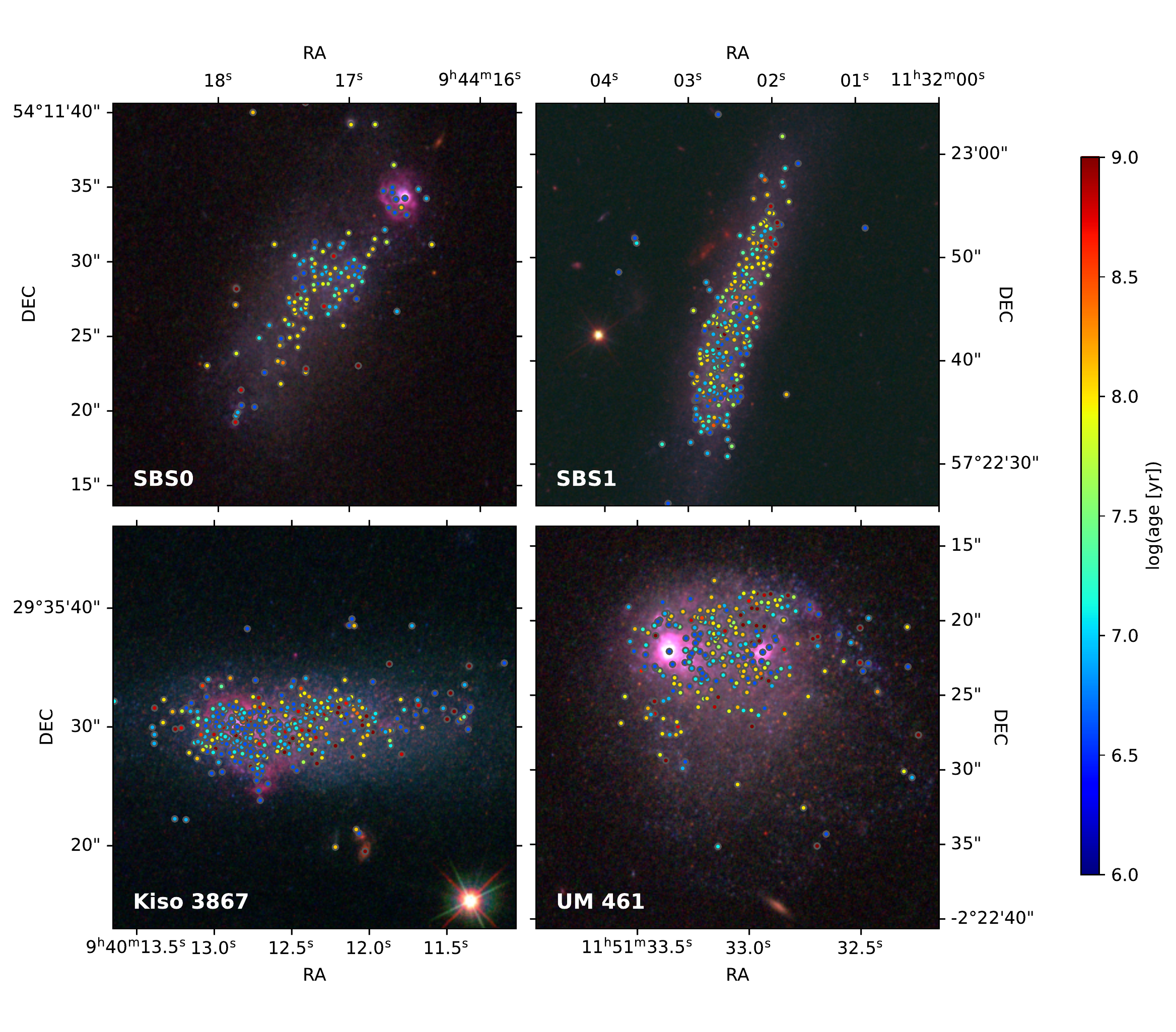}
    %\center{\includegraphics[scale=.75]{composite4-starcluster.png}}
% \includegraphics[width=\columnwidth]{composite4-starclus%ter.png}
    \caption{Color-composite of SBS0, SBS1, Kiso3867, and UM461, overlaid with star cluster ages. The age color scale is shown on the right.}
    \label{fig:cluster_ages}
\end{figure*}

Figure \ref{fig:cluster_ages} shows the spatial distribution of all detected star clusters, color-coded by age, in the 4 galaxies.
The clusters themselves can be seen as bright spots in Figure \ref{color}, where bluer
clusters also correspond to younger ages, although not quantified as in Figure \ref{fig:cluster_ages}.
The results for each galaxy are discussed below.

%fig8

\begin{figure}
    \centering
    \includegraphics[width=\columnwidth]{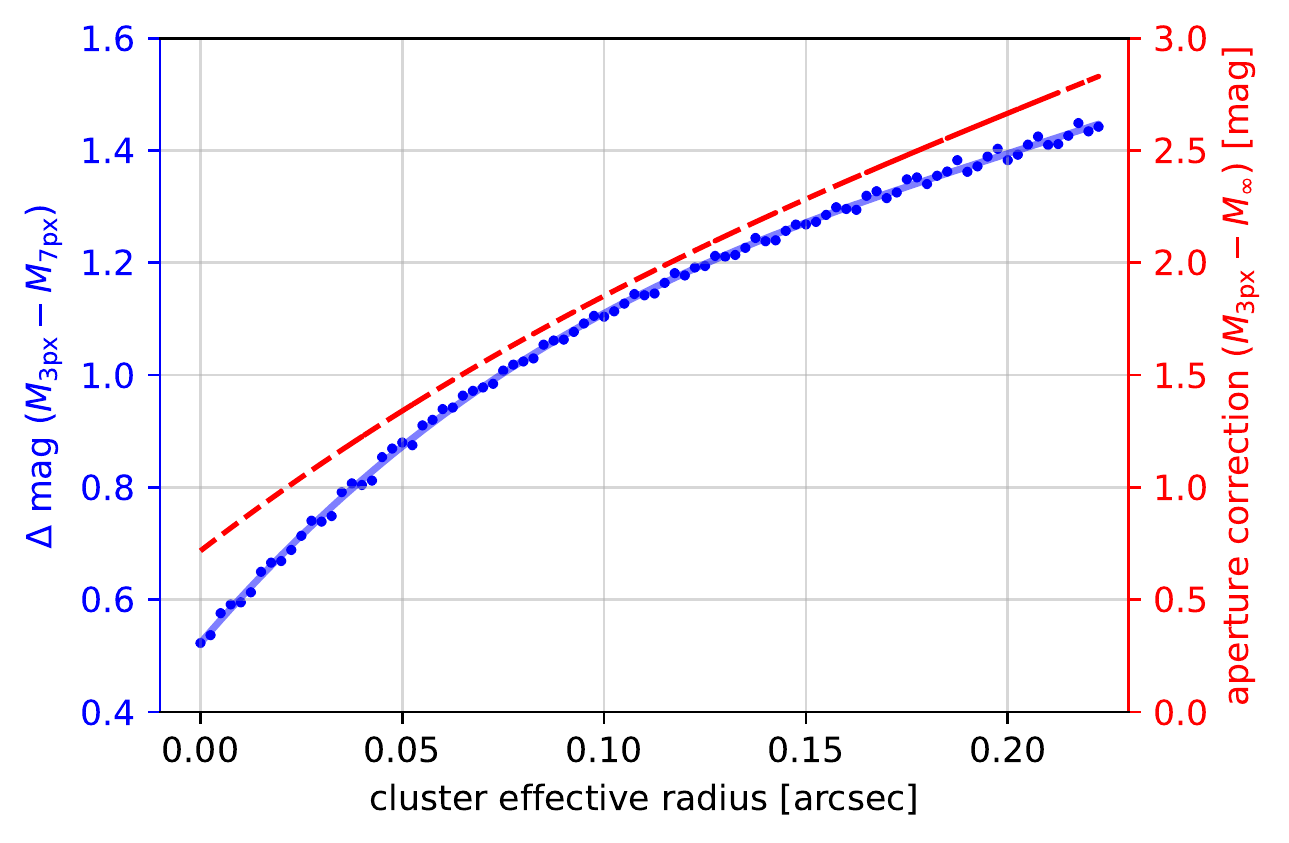}
    \caption{Differential photometry for two apertures varying in size derived for synthetic sources of varying effective radii; Shown in blue, with values along the left y-axis are differential apertures as detailed in the text. Shown in red, with values along the right y-axis, are aperture corrections to correct measurements in the smaller aperture to a total integrated magnitude in an infinitely sized aperture.}
\label{fig:galfitapertures}
\end{figure}

To obtain cluster sizes and from there size-appropriate aperture corrections, we followed the overall strategy presented in \cite{Ryon17}. We generated a large number of model star clusters, convolved with the instrumental PSF, for a range of intrinsic effective radii ranging from 0 to 0.3 arcsec, corresponding to a physical radius from 0 to 16 pc for the nearest galaxy (Kiso~3867). Due to the finite size, we could then derive a correlation between intrinsic size and the differential magnitude between two differently sized apertures; we chose to use the F547M band for this since it is free of nebular emission, and thus the derived star cluster sizes are the least affected by more extended gaseous emission. We chose sizes of 3 and 7 pixels for the two apertures to provide sufficient leverage on the differential photometry, large enough to avoid centering issues, yet small enough to not be significantly affected by nearby sources. From the ensemble of measurements we then derived a spline fit to translate differential photometry directly into intrinsic sizes (first measured in pixels, and converted to physical sizes accounting for the distance to each galaxy). The resulting measurements and derived fits are presented in Figure \ref{fig:galfitapertures}. The same data can also be used to provide size-dependent aperture corrections, using the sizes obtained as described here.

%fig9

\subsection{SBS0}

In SBS0 we find typically young ages $<10$ Myr for sources in and around the main head of the galaxy. The main body of the galaxy shows a variety of cluster ages, ranging from very young to old, with several of the youngest clusters being associated with \ha emission in the main body. We do not observe any significant age trends throughout the main galaxy (besides the overall young ages in the head).

 We measured approximate major axis lengths A$_{major}$ of the tadpoles based on the visible extent of the galaxies on the HST F814W images. The major axis of SBS0 determined by this method is D$_{HST} \approx$ 23~arcsec, corresponding to A$_{major} \approx$ 2.6~kpc for its distance of 24.7~Mpc.

\subsection{SBS1}

SBS1 (D$_{HST} \approx$ 47~arcsec; A$_{major} \approx$ 5.5~kpc) has a single large star-forming site in the south, less dominant than in the other galaxies in this paper, plus a smaller site near the center.The smaller apparent size of the brightest star forming region in SBS1 is primarily due its location in a relatively large galaxy.   There are star clusters scattered across the galaxy, including older clusters between the head and the midsection, with no clear age progression. The H{\sc i} tidal bridge is to the north, on a larger spatial scale than the galaxy, and does not appear to be directly connected to the current star forming patterns.

\subsection{Kiso 3867}
Kiso 3867 (D$_{HST} \approx$ 46~arcsec; A$_{major} \approx$1.5~kpc) has most young clusters near the head of this small dwarf galaxy, but no clear age trend away from the age and into the tail.  Noteworthy are the asymmetric distribution of very young clusters extending away from the midplane of the galaxy, in agreement with the morphology of the nebular emission.

\subsection{UM461}
In UM461 (D$_{HST} \approx$24~arcsec; A$_{major} \approx$ 1.5~kpc), the youngest star clusters are around the two main clumps,
near H$\alpha$ regions, and along the western edge. There are intermediate ages between the clumps,
and
older clusters throughout the galaxy. There is a sprinkling of more young clusters in the north-western part of the galaxy, extending away from the main body of this small system, some of which appear to be associated with faint nebular emission.

\section{Cluster mass and age functions}\label{massfunction}
Cluster mass distribution functions are typically power laws with a slope of around $-1$ on a log-log plot, corresponding to a function $n(M)dM\propto M^{-2}dM$ for equal intervals of mass, M \citep{krumholz19}. Of interest for these galaxies is whether the mass functions have a statistically significant deviation from a power law at high mass, either a decrease as in a Schechter function, or an increase because of high-mass cluster outliers.

The cluster age distribution function is also interesting as this results from a combination of the cluster formation and destruction rates.

In the present study, there are too few clusters to determine a statistically significant probability
distribution function for cluster mass in each galaxy, so a composite function was
made for all the galaxies together by suitable normalization, assuming that this
function is a power law. A composite age distribution function, or cluster decay
rate, was determined as well. As before, we included only clusters with photometric
uncertainties in the F390W band that are  less than 0.2 mag, giving cluster numbers of 38 for SBS0, 96 for SBS1, 135
for Kiso 3867, and 267 for UM461.

\begin{figure}
%\center{\includegraphics[scale=0.5]{kiso_mass-age_4_v2a.jpg}}
\center{\includegraphics[scale=0.5]{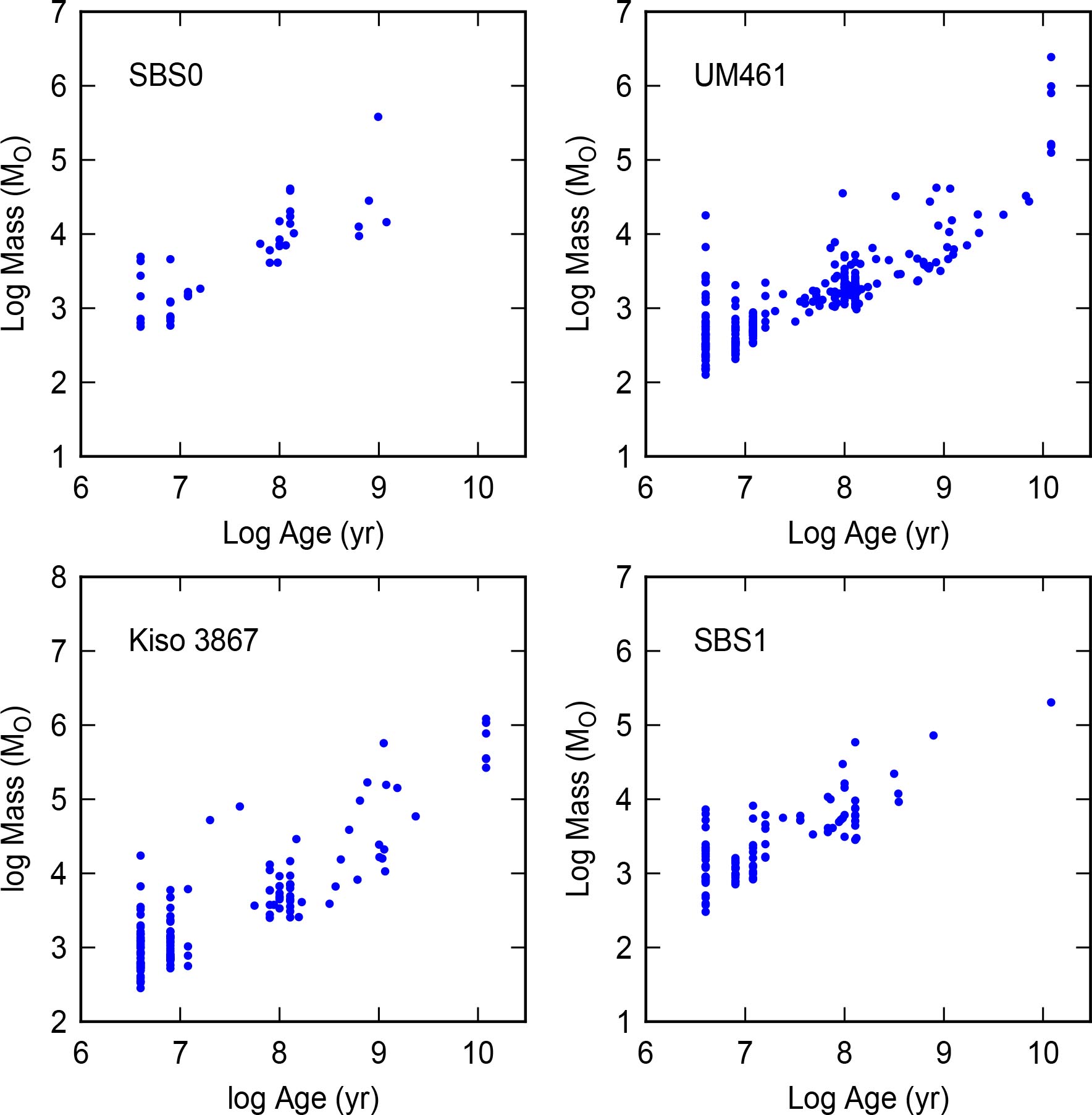}}
\caption{The cluster mass is plotted versus the cluster age for all significant clusters in the four galaxies. The lower limits to the distributions are from detection limits. The cluster numbers decrease vertically according to the cluster mass function, and they decrease to the right according to the cluster age function.}
\label{fig:mass-age}\end{figure}

Figure \ref{fig:mass-age} shows the distribution of cluster mass versus age for
each galaxy. The characteristic lower limit to the mass at each age is from the
limiting apparent magnitude. Concentrations in age around 4 Myr are due to this age being the lowest available age in our model grid, whereas concentrations at 100 Myr and 1 Gyr are due to the variation of (V-I) color with population age.  At $\sim100$ Myr, the (V-I) color changes rapidly due to the appearance of the first red supergiants, followed by a roughly flat color evolution between $10^8$ and $10^9$ years.   When the color changes rapidly,  a large range of color due to measurement uncertainties corresponds to a small range of age, concentrating the clusters at that age. In practice, due to the limited SED coverage of our observations, the SED fitting likely favors an age at either the young or the old limits of this color-plateau.

%fig10

To make a composite mass-age diagram, we multiply all the cluster masses by
$(10/D)^2$ for galaxy distance $D$ in Mpc. This increases the masses for
closer galaxies and decreases them for distant galaxies, making the detection
limit the same for each, as if they were all at 10 Mpc.

\begin{figure}
%\center{\includegraphics[scale=.6]{kiso_mass-age_4_v2.jpg}}
\center{\includegraphics[scale=1.]{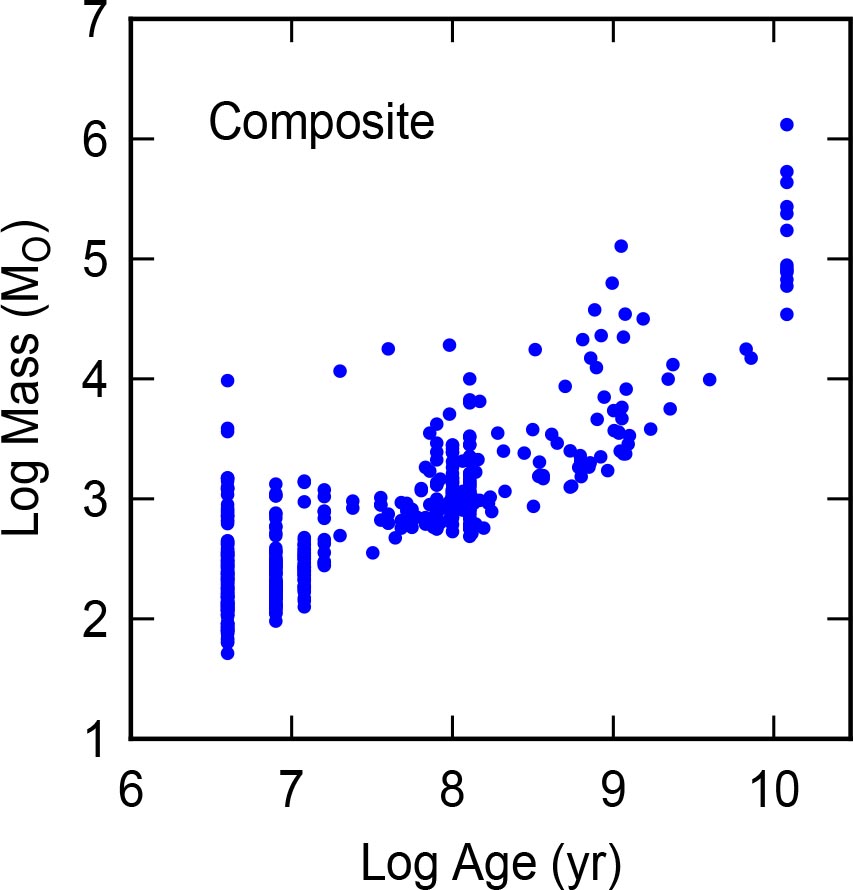}}
\caption{The cluster masses are plotted versus age for all the clusters combined from the four galaxies. Cluster masses were normalized to a common distance of 10 Mpc.}
\label{fig:mass-age2}\end{figure}

After this, the lower
limits to the mass-age distributions are all the same and they can be stacked into a composite diagram. This composite is shown in Figure \ref{fig:mass-age2}.  At each age, the
vertical distribution of points is the cluster mass function. At each mass, the horizontal distribution of points is the age distribution, which for a constant formation rate is the survival distribution function, i.e., the number of clusters of a particular mass that survive as a function of age. Typically the vertical density of points tapers off like $\sim1/M$, which means the mass function is $dn/dM\sim M^{-2}$, and the horizontal density of points is approximately constant or increasing slightly, meaning the cluster number decreases like $1/$age or shallower \citep{whitmore07}.

%fig11

\begin{figure}
%\center{\includegraphics[scale=0.5]{kiso_mass_function4_v2.jpg}}
\center{\includegraphics[scale=0.52]{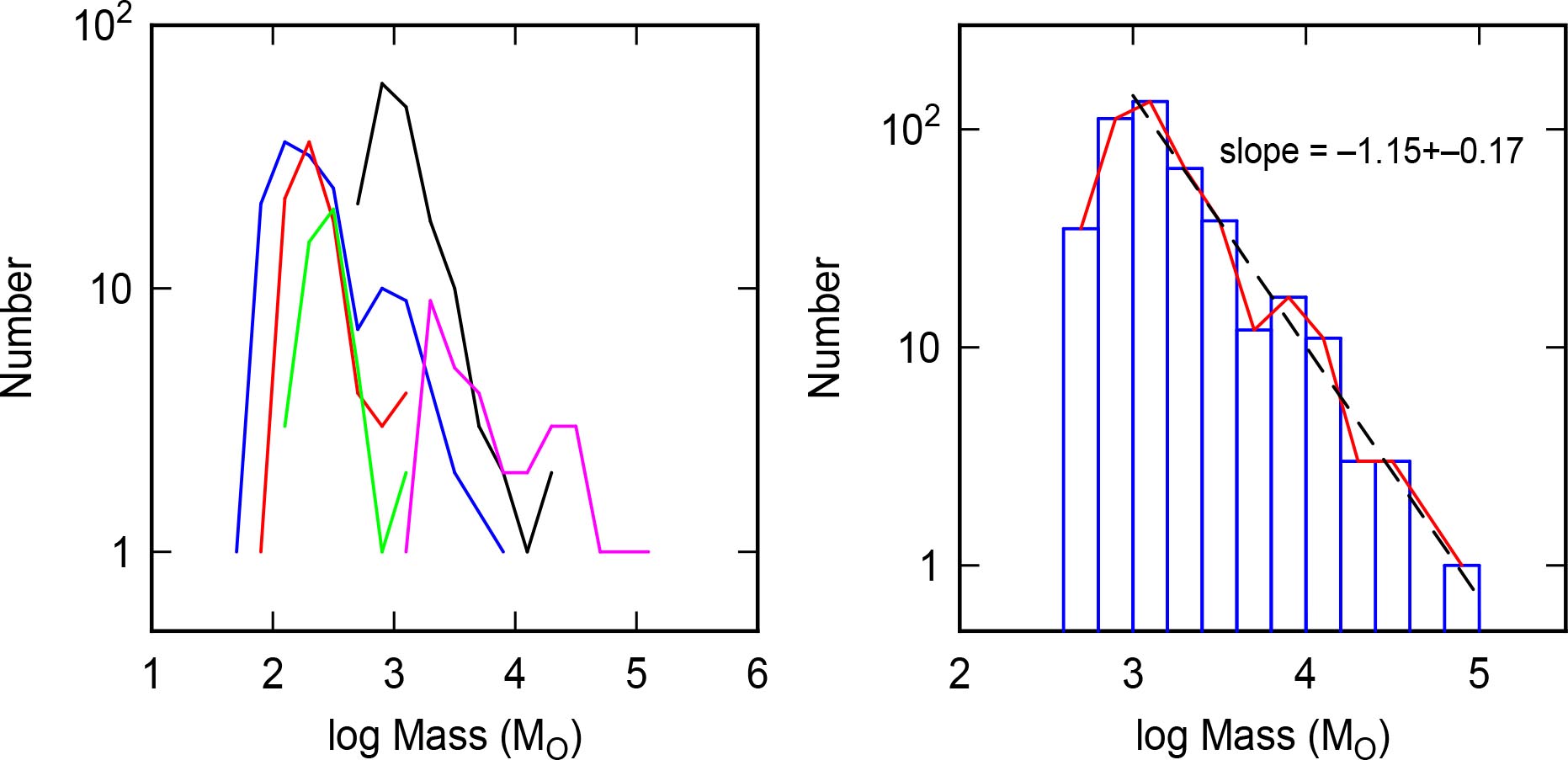}}
\caption{Mass distribution function for all clusters combined in the four tadpole galaxies. Left: After a first normalization to a common distance of 10 Myr, the cluster mass distributions are shown in 5 age bins, i.e., three around 4, 8, and 12 Myr (blue, red, green), and the ranges given by $\log({\rm age})=7.6$ to 8.3 (black) and 8.8 to 9.2 (magenta).  There is a slight shift to the right with increasing age, reflecting the slope of the detection limit. Right: The stacked composite cluster mass function obtained after a second normalization that shifts all the lower limits from detection losses in each age bin to the lower limit at $\log({\rm age})=7.6$ to 8.3.  These normalizations conserve the slope of the cluster mass function if it is a power law. The result in the figure is a power law, confirming the assumption, and the slope of the composite power law, $-1.15\pm-0.17$, is the best representation for the composite function. There are no outliers at high mass and there is no indication of a fall-off as in a Schechter function.}
\label{fig:kiso_mass_function3_v2}\end{figure}

The cluster mass function for distance-normalized cluster masses was determined in
two ways, first for a few narrow age intervals and then by stacking these
intervals with appropriate vertical shifts. The age intervals were taken to be the
three discrete ages at small age, namely, 4, 8, and 12 Myr, and the two other
intervals with age concentrations, $\log({\rm age})=$ 7.6 to 8.3, and $\log({\rm age})=$8.8 to
9.2, for ages in years.  These five intervals give five mass functions, which are
each fairly noisy because of the small numbers of clusters in each. The left-hand panel of
Figure \ref{fig:kiso_mass_function3_v2} shows the resulting composite
mass functions in the five age intervals, where each one combines all the galaxies by the distance normalization. There is a systematic shift toward higher masses for older clusters from the rising lower limit to the mass-age plot in Figure \ref{fig:mass-age}.

The age intervals can be combined by shifting the masses of the younger clusters
upward by an appropriate amount. Stacking the result essentially assumes the mass
function is a power law, but it does not assume any particular power. To do this stacking,
we multiply all the cluster masses in each age interval by the ratio of the
minimum cluster mass in the $\log({\rm age})=7.6$ to 8.3 interval to the minimum in
that interval.  This makes all the minimum masses have the same minimum as the 7.6
to 9.3 log-age interval. All five shifted mass functions
are then stacked to make one composite mass function. The result is shown on the right in
Figure \ref{fig:kiso_mass_function3_v2}.

%fig12

Figure \ref{fig:kiso_mass_function3_v2}-right shows a smooth power law mass function with a slope of $-1.15\pm0.17$
(90\% confidence interval by the student-t distribution). This agrees with the
idealized case resulting from wide-spread hierarchical structure in star
formation, where the slope on a log-log plot would be -1 \citep{elmegreen97}. There are no
outlier masses beyond the power law at the high end, and no drop-off there either, as in a Schechter function \citep[see also][]{mok20}.

\begin{figure}
%\center{\includegraphics[scale=0.5]{kiso_age_function4_v2.jpg}}
\center{\includegraphics[scale=0.52]{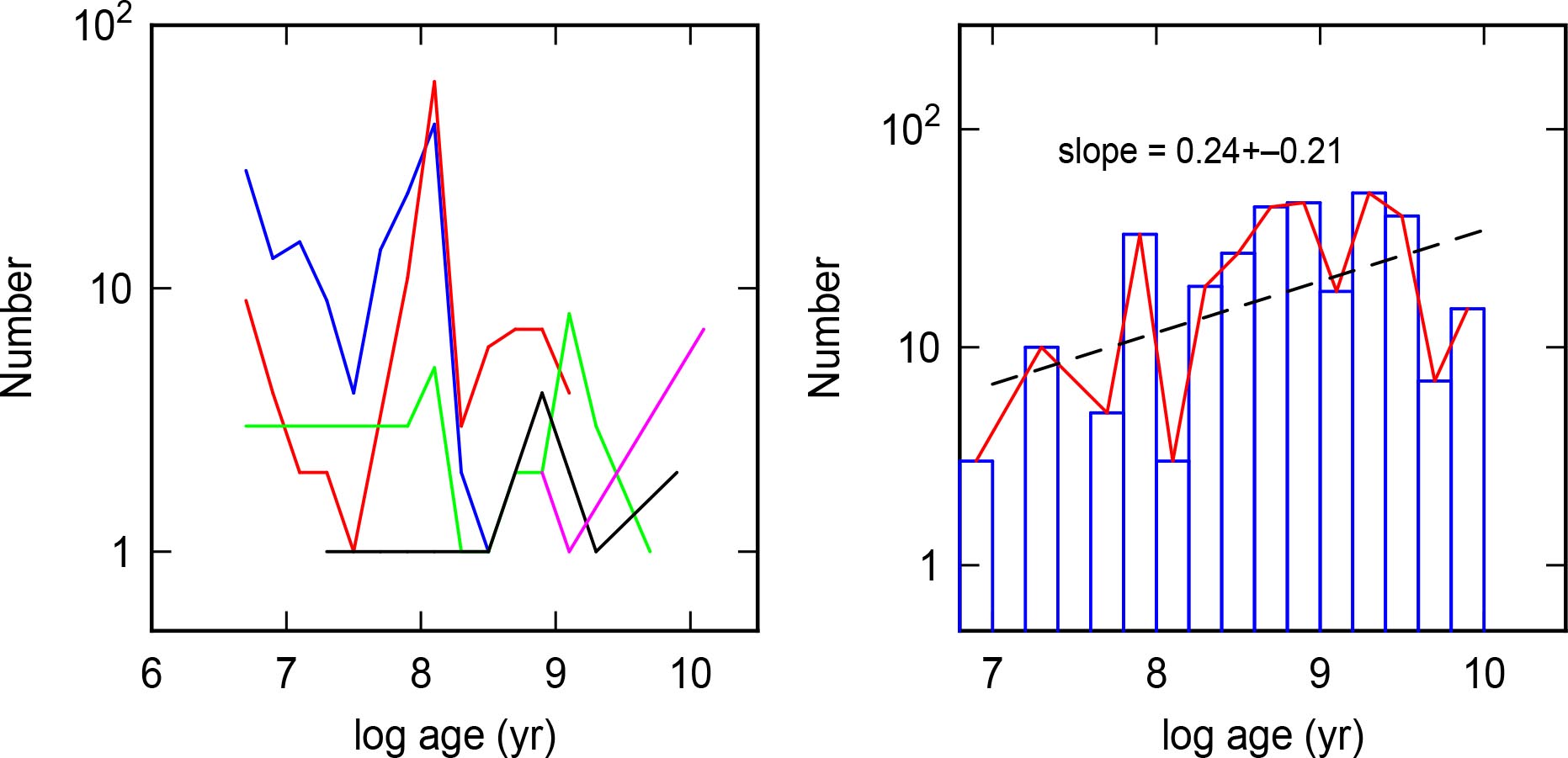}}
\caption{Age distribution function for all clusters combined in the four tadpole galaxies. Left: After a first normalization to a common distance of 10 Myr, the cluster age distributions are shown in 5 mass bins, i.e.,
$\log(mass)=2.5-3$, $3-3.5$, $3.5-4$, $4-4.5$ and $4.5-5$ in solar masses (with curve colors blue, red, green black and magenta, respectively).
There is a slight shift to the right with increasing mass, reflecting the slope of the detection limit.
Right: The stacked composite cluster age function obtained after a second normalization that shifts all the upper limits from detection losses in each age bin to the upper limit at $\log({\rm age})=4$ to 4.5.  These normalizations conserve the slope of the cluster age function if it is a power law.
}
\label{fig:kiso_age_function3_v2}\end{figure}

The average age distribution can be found in a similar way, assuming the actual
distribution is a power law in age, by taking the age distributions in narrow
distance-normalized mass windows and shifting them all to have the same maximum
value, again following the detection limit in the mass-age diagram. The distance-normalized mass intervals are given by log (Mass) $(M_{\odot})=2.5-3$, $3-3.5$, $3.5-4$, $4-4.5$ and $4.5-5$. The result is shown
in Figure \ref{fig:kiso_age_function3_v2}.  The slope of the distribution is $0.24\pm0.21$ (90\% confidence
interval by the student t-distribution).  This slope implies that the number of clusters
decays with time as $dn(t)/dt\propto t^{-0.76\pm0.21}$.

%fig13

%fig14
\begin{figure*}
\center{\includegraphics[scale=0.6]{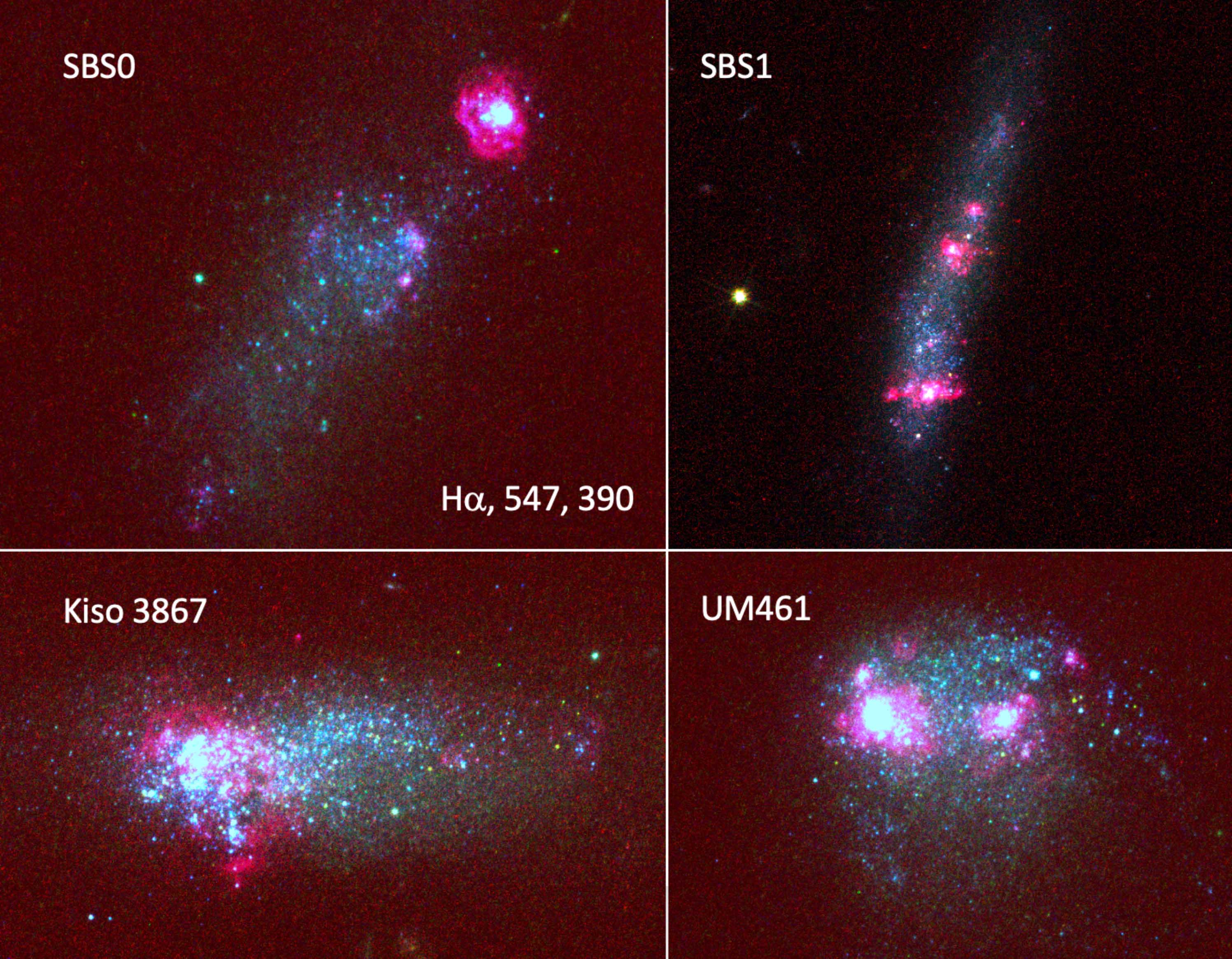}}
\caption{Color composite images using F390W for blue, F547M for green, and F657N (not continuum subtracted) for red.}\label{fig:colorcomposite}\end{figure*}

\vspace{0.1in}

\vspace{0.1in}
\section{H$\alpha$ Emission}\label{sect:Ha}
\subsection{Cluster positions and H$\alpha$}
Figure \ref{fig:colorcomposite} shows a color composite image using F390W for blue, F547M for green,
and F657N including the H$\alpha$ emission for red.
Overall we find good agreement between location of young clusters and \ha regions, but often H$\alpha$ emission is noticeably more extended, which is not unexpected in these comparably low density environments. We do not find age gradients of the clusters inside the complexes nor in the galaxies as a whole.
We also notice several young star clusters with no significant \ha emission, with no obvious explanation. Possible reasons could be: low gas density and thus low \ha surface brightness below our detection threshold; many of the clusters are also very low mass, so overall \ha luminosity is expected to be low as well; lastly, we can not rule out that at least some fraction of the ionizing radiation is leaking out of the immediate vicinity of the cluster, potentially contributing to \ha luminosity elsewhere or even leaving the galaxy altogether.

It therefore appears that the intense star forming regions in our sample galaxies are likely special events, but are not isolated, as shown by the ``halos'' of surrounding less massive young star clusters. The event that produced the tadpole ``heads'' involved the production of large scale massive clouds, but was also associated with more distributed star formation activity extending throughout the larger volume of the galaxy. Again it is not clear if these are associated -- with some event triggering both distributed and concentrated star formation -- or distinct phenomena where the particular tadpole features are triggered separately and are merely embedded into the pre-existing star formation activity throughout the galaxy.
 Clustered but disorganized age patterns are typical of low mass irregular and related galaxies.

\subsection{\ha measurements}
In order to measure the total H$\alpha$ luminosity of each of the four galaxies, we integrated the total counts in the continuum-subtracted H$\alpha$ image over the region dominated by the line emission. In doing so, care must be taken to include all pixels with line emission, and at the same time exclude adjacent pixels with no emission that would only add to the noise in the measurement. A simple, effective way to achieve this is to convolve first the image with a Gaussian, then build a mask by setting to 1 all pixels above a given threshold (and 0 the pixels below such threshold) in the convolved image. The continuum subtracted H$\alpha$ image is then multiplied by this mask, and the total counts computed. Some contaminant objects in the field, but clearly not associated to the galaxy, were masked out by hand.

We found that a convolution with a Gaussian with ${\rm FWHM} = 12$ pixels and a threshold corresponding to a flux of $19\times 10^{-17}$ ergs s$^{-1}$ cm$^{-2}$ arcsec$^{-2}$ (approximately the rms of the background) worked best for our objects.

The conversion factor from counts to flux was computing by using the formula:

$$\int F_\lambda d\lambda = C \frac{hc}{\lambda_0} \frac{1}{T(\lambda_0)} \frac{1}{A_\mathrm{geo}}$$

where $C$ is the count rate, $\lambda_0$ is the wavelength of the redshifted H$\alpha$ line, $T(\lambda_0)$ is the transmission value of the throughput of the F657N filter at $\lambda_0$, and $A_\mathrm{geo}$ the collecting area of the telescope. Developing the above formula we obtain:

\begin{displaymath}
\int F_\lambda d\lambda = 2.65\times10^{-16}\,\times C
\end{displaymath}
(in ergs s$^{-1}$  cm$^{-2}$).

The conversion factors computed this way agree very well with the one computed by the product $\textrm{PHOTFLAM} * W_\mathrm{eff}$, where PHOTFLAM is the value recorded in the H$\alpha$ image header, defined as "the flux of a source with constant flux per unit wavelength (in ergs s$^{-1}$  c$^{-2}$ $\lambda^{-1}$) which produces a count rate of 1 DN per second", and $W_\textrm{eff}$ is the effective width of the F657N filter. Note however that this latter method does not take into account the variation of the filter transmission curve as a function of $\lambda$.

In all four galaxies, the main H$\alpha$ clumps have been identified as those H$\alpha$ emitting regions inside which the flux, in each pixel, is above 75 $\times 10^{-17}$ ergs s$^{-1}$ cm$^{-2}$ arcsec$^{-2}$, and the total H$\alpha$ flux measured. The central parts of the bigger clumps were identified by setting a threshold 2 mag higher.

For both the total and the clump H$\alpha$ emission the corresponding star formation rate was computed from the following formula \citep{kennicutt12}:
$$\log({\rm SFR})= -5.84 +\log(F(H\alpha)) + 2*\log(D/15) $$
where SFR is in ${M}_\odot$ yr$^{-1}$, $F(H\alpha)$ is in $10^{-17}$ ergs s$^{-1}$ cm$^{-2}$, and the distance $D$ is in Mpc.

Table~\ref{tabFHa} lists, for each galaxy and its clumps, the total H$\alpha$ flux, the integration area, and the SFR.

%fig15
\begin{figure*}
\mbox{
\centerline{
\hspace*{0.0cm}\subfigure{\includegraphics[width=0.5\textwidth]{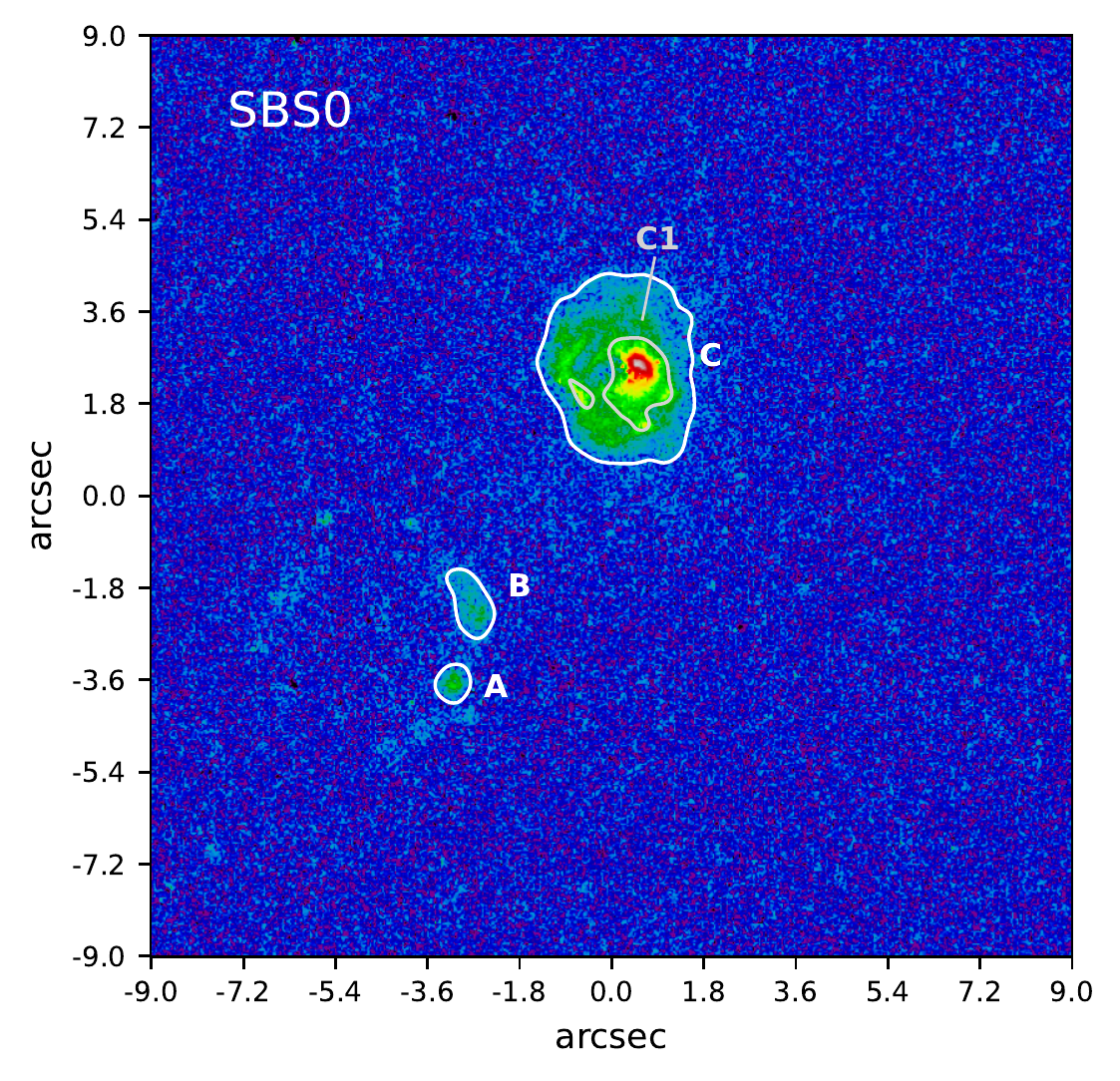}}
\hspace*{0.0cm}\subfigure{\includegraphics[width=0.5\textwidth]{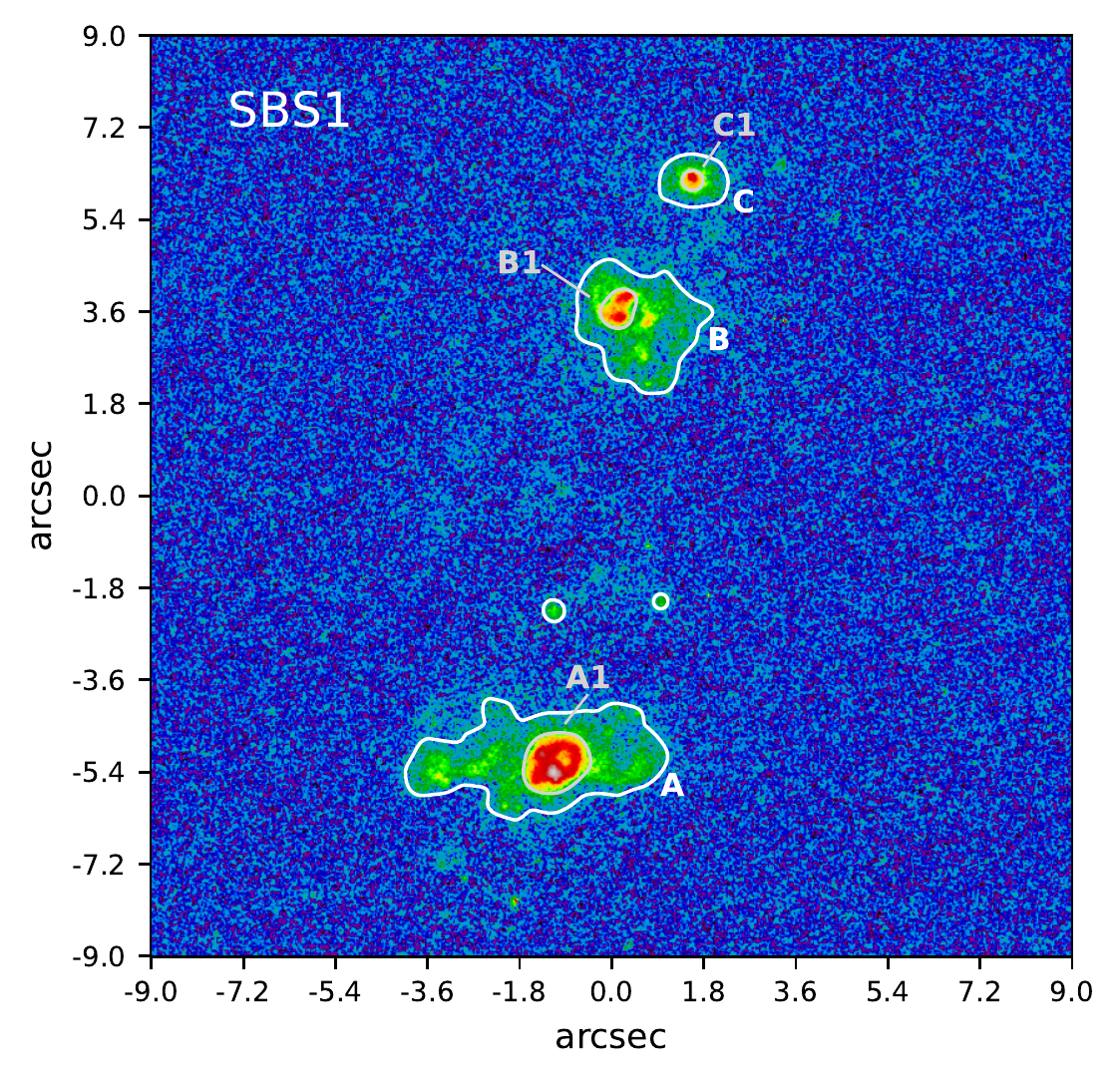}}
}}
\mbox{
\centerline{
\hspace*{0.0cm}\subfigure{\includegraphics[width=0.5\textwidth]{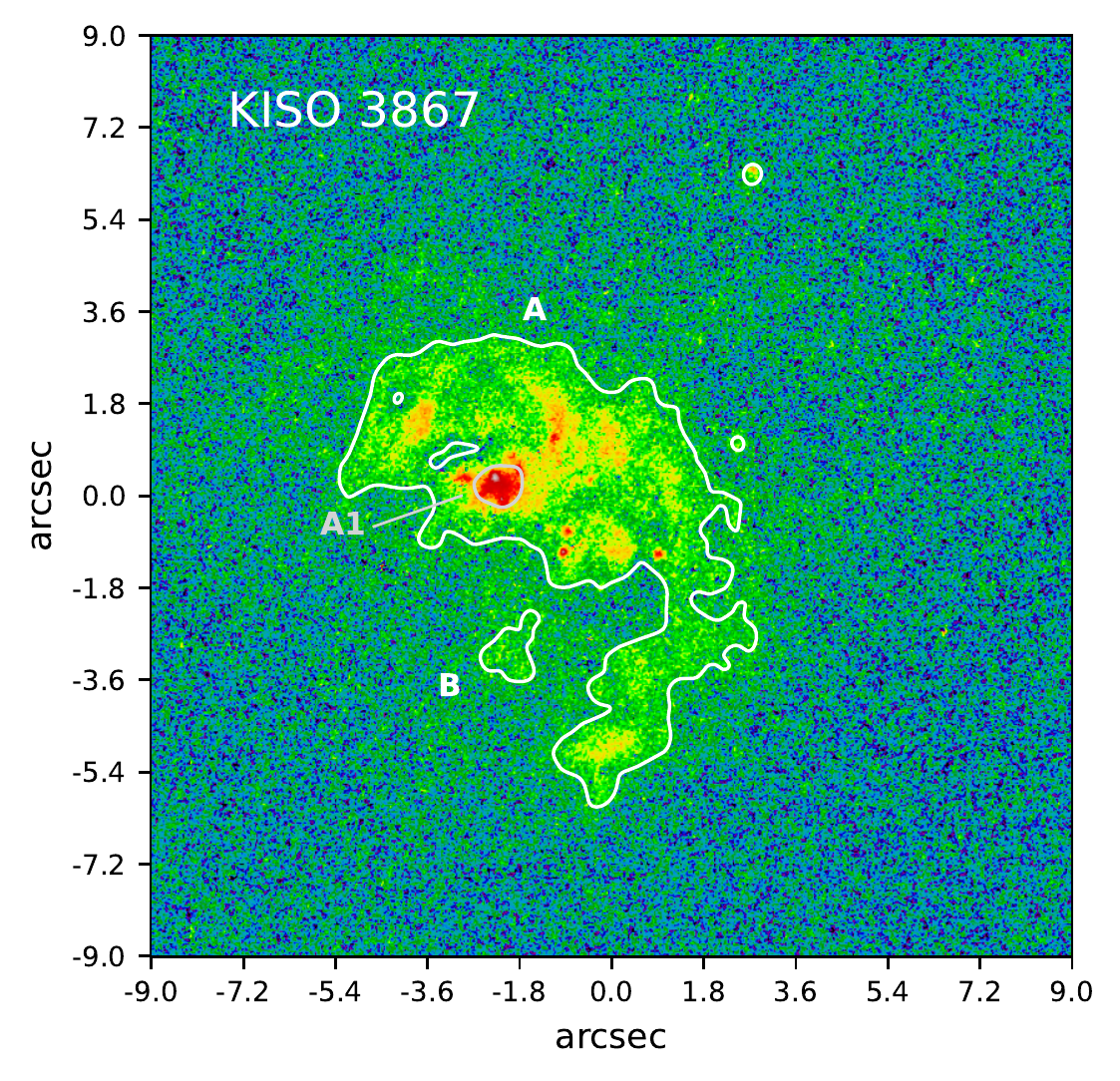}}
\hspace*{0.0cm}\subfigure{\includegraphics[width=0.5\textwidth]{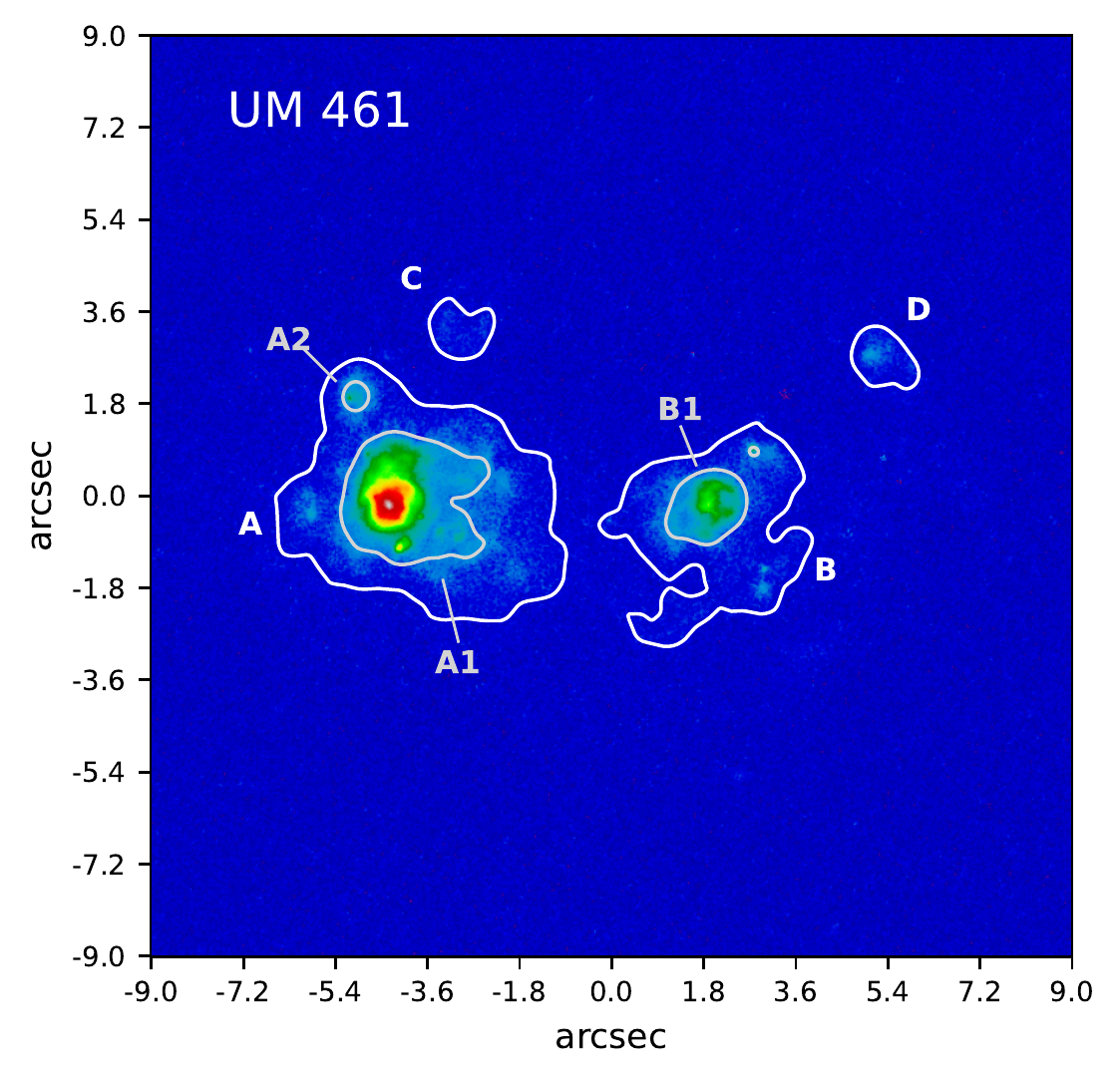}}
}}
\caption{The main clumps identified and labeled on a H$\alpha$ map of the galaxies, shown in logarithmic scale. The labels correspond to the clumps listed in Table~\ref{tabFHa}. The white contours correspond to a flux level of 75 $10^{-17}$ ergs s$^{-1}$ cm$^{-2}$ arcsec$^{-2}$, while gray contours are 2 mag brighter.}
\label{Fig:Haclumps}
\end{figure*}

The uncertainty on the computed H$\alpha$ flux inside a given region is largely due to the accuracy of the continuum subtraction in the H$\alpha$ image. To estimate this uncertainty, we computed two other
continuum-subtracted H$\alpha$ images, the first by increasing by $\pm20\%$ the coefficient $a$ in equation \ref{Eq:contHa}, and decreasing $b$ by the same amount, the second the other way around. The H$\alpha$ fluxes were then computed on these two images, inside the same star-forming knots, and compared to the correct values to derive an average difference, expressed in percentage.

The error on the integrated flux of the H$\alpha$ emission depends more on the exact value of the threshold used to compute it than on the accuracy of the continuum subtraction. We thus recomputed the integrated flux by changing by $\pm 20\%$ the adopted threshold, and estimated the corresponding uncertainty (in percentage) as half the difference between these two values divided by the H$\alpha$ flux.

\begin{figure}
%\center{\includegraphics[scale=.45]{clumpmassvHaSFR.pdf}}
\center{\includegraphics[scale=.45]{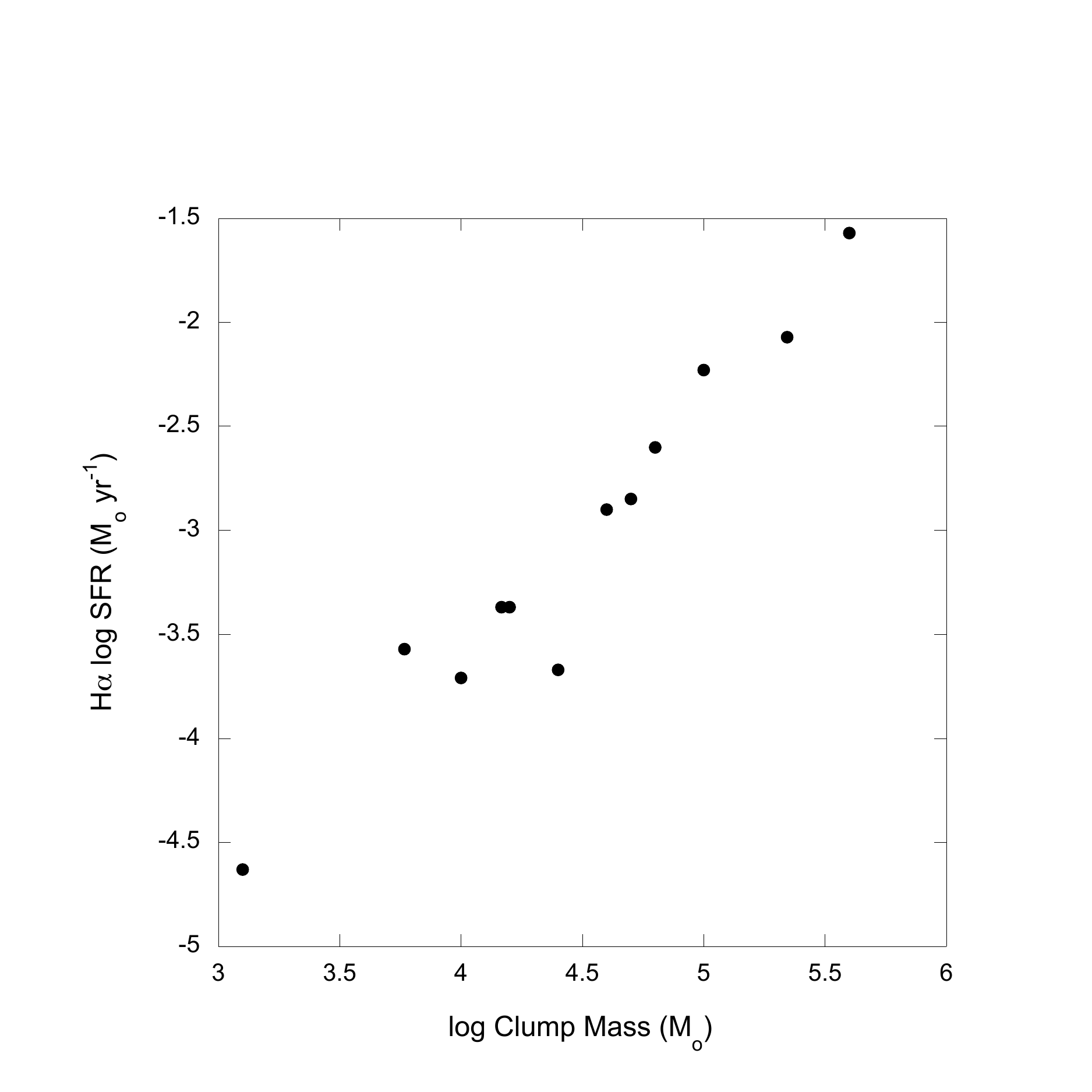}}
\caption{The star formation rates based on H$\alpha$ emission are plotted versus the corresponding large-scale clump masses in the four galaxies.}
\label{fig:Havclump}\end{figure}

Many of the H$\alpha$ clumps correspond with the large-scale star-forming clumps described in Sect.~\ref{sect:clump} and shown in Figure \ref{contour}. The star formation rates based on the H$\alpha$ measurements are plotted versus the clump masses in Figure \ref{fig:Havclump}, where a tight correlation is seen. The average ratio of the clump mass to the  H$\alpha$ SFR is $31\pm1.8$ Myr, which is one possible measure of the mean lifetime of the regions. This lifetime is longer than the photometric ages given in Table \ref{tab2}, which were used to derive the SFRs there. The average photometric age of the clumps is 4.7 Myr for all galaxies if the extinction is taken to be a free parameter, increasing to 16.4 Myr if $A_V=0.1$ is forced. These small values compared to 31 Myr are a hint that some H$\alpha$ flux may be missing, perhaps a factor of 2 or more for the young stellar clumps.
In other words, the SFR from H$\alpha$ is slightly smaller than the SFR from SED fitting. The next section discusses the possibility of missing H$\alpha$
from a consideration of star cluster SED fits.

%fig16

\begin{deluxetable}{lcccc}
\tabletypesize{\scriptsize}\tablecolumns{4}
\tablewidth{0pt} \tablecaption{Fluxes, areas, and SFRs for the H$\alpha$-emitting clumps\label{tabFHa}}

\tablehead{
\colhead{Clump}&
\colhead{H$\alpha$ flux} &
\colhead{Uncert.} &
\colhead{Area} &
\colhead{log SFR}
\\
\colhead{}&
\colhead{ergs s$^{-1}$ cm$^{-2}$ 10$^{-17}$} &
\colhead{percent} &
\colhead{(arcsec$^2$)}&
\colhead{(M$_{\odot}$ yr$^{-1}$)}
}
\startdata
SBS0: 	    &		      &               &               &       	   \\
total       &   $4873$    & $5.9\%$       & $31.26$       & $-1.72$    \\
A    	    &	$69$      & $2.2\%$       &	$0.40$        &	$-3.57$    \\
B           &   $109$     & $3.0\%$       &	$0.80$        &	$-3.37$    \\
C           &   $3941$    & $< 1\%$		  &	$9.15$        &	$-1.81$    \\
C1          &   $2192$    & $< 1\%$       &	$1.65$        &	$-2.07$    \\[4pt]
SBS1:       &		      &				  &	              &            \\
total       &   $5091$    & $9.1\%$       & $47.35$       & $-1.72$    \\
A	        &   $2551$	  & $1.6\%$	      & $7.63$ 	      & $-2.02$    \\
A1	        &   $1575$    & $< 1\%$       & $1.24$	      & $-2.23$    \\
B	        &   $989$     & $2.1\%$       &	$4.50$	      &	$-2.43$    \\
B1	        &   $330$     & $1.2\%$       &	$0.41$	      &	$-2.90$    \\
C           &   $278$  	  & $2.2\%$       &	$1.13$        &	$-2.98$    \\
C1     	    &   $112$     & $1.1\%$       & $0.13$        &	$-3.37$    \\[4pt]
Kiso 3867:	&			  &		          &	              &	           \\
total       &   $7777$    & $6.7\%$       & $106.99$      & $-2.65$    \\
A	        &   $4925$	  & $2.5\%$       &	$31.91$       &	$-2.85$	   \\
A1	        &   $467$     & $1.7\%$       &	$0.61$        & $-3.87$	   \\
B	        &   $81$      & $2.6\%$       &	$0.89$        & $-4.63$	   \\[4pt]
UM461:	    &		      &	              &				  &            \\
total       &   $31281$   & $1.2\%$       & $84.44$       & $-1.43$    \\
A	        &	$25517$   & $< 1\%$       &	$19.29$       &	$-1.52$    \\
A1	        &	$22793$   & $< 1\%$	      &	$5.64$ 		  &	$-1.57$    \\
A2	        &	$162$ 	  & $1.1\%$       &	$0.22$        &	$-3.71$    \\
B	        &	$3425$    & $1.3\%$       & $10.30$       &	$-2.39$    \\
B1	        &	$2113$    & $< 1\%$  	  &	$1.76$        &	$-2.60$    \\
C	        &	$128$     & $2.1\%$ 	  &	$1.07$        &	$-3.82$    \\
D	        &	$180$     & $1.4\%$ 	  &	$1.15$        &	$-3.67$    \\
\enddata
\tablecomments{The uncertainties on the fluxes of the star-forming knots is computed by subtracting various continua, while the uncertainty on the total flux was derived by changing the flux threshold for integration. See text for details.}
\end{deluxetable}

\subsection{Comparison between observed \ha flux and predictions from star cluster SED fits}

Combining theoretical expectations from the stellar population models and the observed best-fit parameters derived for the star clusters in each galaxy we can make an estimate on the expected flux of H-ionizing photons of the entire cluster population given the age and stellar mass of each star cluster. The results for this comparison are presented in Table~\ref{tab:lyc}. NLyC(H$\alpha$) is the observed emission rate of Lyman continuum photons in units of $10^{50}$ s$^{-1}$ that is required to power the integrated H$\alpha$ emission from the galaxy, assuming Case-B and a standard conversion ratio of 2.2 between number of ionizing and H$\alpha$ photons \cite{osterbrock06}.
NLyC(SED) is the theoretical summed emission rate from the clusters given their stellar populations, which were derived from the SED fits to each one (see Sect.~\ref{sect:cluster}), before aperture correction.  NLyC(SED-Corr) is the Ly-C emission rate obtained from SED fits corrected with the aperture correction appropriate for each cluster given its size (also see Sect.~\ref{sect:cluster}). As such it presents the intrinsic ionizing photon production rate summed over all discrete, detected clusters, but before any attenuation due to possible surrounding dust; since dust attenuation is not known {\it a priori} and may be different from the reddening derived for the cluster light due to radiation transfer effects, these SED-corrected values thus present an upper limit to the actual, detectable ionizing flux.

\begin{table}[]
%\centering
\begin{center}

\caption{Observed and Calculated Ionizing Fluxes}
\begin{tabular}{cccc}
Galaxy     &NLyC& NLyC& NLyC\\
&from $H\alpha$&SED&SED-Corr\\\hline
SBS0       &   $8.3 \pm 0.4$    & 3.8           &  33 \\
SBS1       &   $8.2 \pm 0.3$    & 9.4           & 210 \\
Kiso 3867  &   $1.1 \pm 0.2$    & 1.6           &  21 \\
UM461      &  $16.4 \pm 0.08$   & 3.7           &  38 \\
\end{tabular}
\label{tab:lyc}
\end{center}
\tablecomments{The galactic Lyman continuum flux in units of $10^{50}$~s$^{-1}$ from three determinations. The first is the LyC flux needed to ionize the observed $H\alpha$. The second is predicted from the cluster SEDs based on modeling the small aperture fluxes used to derive cluster colors. The third uses the same procedure with large apertures that contain the full flux. Since large apertures are likely to contain additional light sources, this third SED-corrected value is an upper bound to the NLyC predicted by the cluster simple stellar population models.}
\end{table}

For SBS1 and Kiso 3867, the required Lyc flux from the observed H$\alpha$ is smaller than the theoretical Lyc flux from the SED fits to the clusters. Considering also the larger regions around the clusters, the SED fits would predict more than 20 times larger H$\alpha$ fluxes that what is observed.

A feature of all of these galaxies, and especially SBS1 and Kiso 3867, is the large number of compact stellar groups and clusters. Kiso~3867 is the nearest system in our sample, which may partially account for the large number of resolved stellar clusters and groups. The ongoing interaction between SBS1 and its companion likely affected its evolution. These compact regions can readily supply a significant fraction and possibly all of the ionizing Lyman continuum flux from stars. This observation suggests first that $\Gamma$, the fraction of young stars in clusters, may be high and at minimum $>$10\%. This result is interesting in terms of the possibility that these galaxies may be forming an unusually high fraction of their stars in bound clusters \citep[e.g.,][]{adamo15, krumholz19}.

The large predicted H$\alpha$ fluxes compared to the observed values, particularly for SBS1 and Kiso 3867, raise the possibility that some of the expected LyC photons are being absorbed by dust or escaping the observed HII regions and diffuse H$\alpha$.
Low metallicity galaxies like these typically have more LyC escape \citep{ramambason22}.
Deeper images will be required to determine if escaping LyC is ionizing an extended halo around the galaxies, or is possibly escaping the galaxies altogether.

There are also systematic uncertainties in the LyC flux derivations. For the  observations they include: a) uncertain continuum removal from the F657N filter to derive the H$\alpha$ fluxes; b) dust reddening corrections for H$\alpha$; c) conversion of H$\alpha$ flux to number of ionizing photons, and d) undetected H$\alpha$ emission from possible heavy obscuration. There are also uncertainties in the SED estimates, such as: a) the star formation history; b) the IMF, especially the upper limit; c) the broad band photometry, which uses only three filters, and d) dust reddening corrections for each cluster SED.

\vspace{0.1in}
\section{Conclusions}
\label{sect:conc}

Multiwavelength imaging of four tadpole galaxies with HST WFC3 and ACS reveal a pattern similar to that previously found for Kiso~5639. The three largest galaxies contain one or two giant stellar complexes with masses of $10^5\;M_\odot$ or more and ages of a few million years. The smallest galaxy, Kiso 3867, has a maximum complex mass of $\sim5\times10^4\;M_\odot$ and a similarly young age. There are typically smaller complexes nearby that are also young.

The complexes are accompanied by large populations of compact young star clusters. The mass distribution function for the combined clusters was determined by a double-normalization technique, which assumes only that the intrinsic function is a power law. In this technique, the cluster masses were first scaled as if all the galaxies were at the same distance. Then the cluster masses in five age bins were scaled a second time to make their distributions have the same lower mass cutoff from sensitivity limits. The composite mass function determined in this way was indeed a power law, confirming the initial assumption, and the slope on a log-log plot was $-1.15\pm0.17$, similar to the expected value of $-1$ for hierarchical distributions. There were no significant deviations from this power law at high mass, i.e., no outliers with masses higher than the extrapolation of the function, and no suggestion of an upper mass cutoff.

The composite age distribution function was determined in a similar way, although there is no {\it a priori} reason for this to be a power law. Combining all masses between $10^{2.5}\;M_\odot$ and $10^5\;M_\odot$, the result showed a slight increase in cluster counts with log (Age) $t$, as $t^{0.24\pm0.21}$, which corresponds to a loss of clusters over time as $t^{-0.76\pm0.21}$.

The tadpole galaxies in our sample share the common feature of reduced metallicity in the most prominent star-forming complex as compared with its surroundings, as determined in previous papers. This suggests that recent accretion of low-metallicity gas triggered the main starburst \citep[e.g.,][]{jorge13}.  The gas could accrete in several ways.
The off-center locations of the main complexes could be the result of tadpole motions relative to a surrounding lower-metallicity medium if the head is more compressed than the rest of the disk by ram pressure. Alternatively, the metal-poor gas could be accreted from a gaseous halo or cosmic filament with non-zero specific angular momentum.
The quiescent kinematics of most tadpoles appears to rule out fueling from minor mergers. Aside from this accretion, these tadpole galaxies appear to be typical blue compact dwarfs.

The summed theoretical Lyman continuum (LyC) emission from the clusters is comparable to the LyC needed to excite the observed H$\alpha$, but the summed LyC emission including regions around the clusters is more than 20x larger. This excess LyC suggests a loss of LyC photons, which could be explained by dust absorption although these galaxies have low metallicities. Alternatively there could be additional H$\alpha$ emission in a halo that we do not see, or some escape of LyC photons from the galaxy.

{\it Acknowledgments:}

The imaging data presented in this paper were obtained from the Mikulski Archive for Space Telescopes (MAST) at the Space Telescope Science Institute. The specific observations analyzed can be accessed via \dataset[10.17909/kjp2-sh47]{https://doi.org/10.17909/kjp2-sh47}.
Based on observations with the NASA/ESA Hubble Space Telescope obtained at the Space Telescope Science Institute, which is operated by the Association of Universities for Research in Astronomy, Incorporated, under NASA contract NAS5-26555. Support for HST-GO-15860 was provided through a grant from NASA/STScI. JSA, CMT, and NC acknowledge support from the Spanish Ministry of Science and Innovation, project  PID2019-107408GB-C43 (ESTALLIDOS), and from Gobierno de Canarias through EU FEDER funding, project PID2020010050.
This research has made use of the NASA/IPAC Extragalactic Database (NED) which is operated by
the Jet Propulsion Laboratory, California Institute of Technology, under contract with the
National Aeronautics and Space Administration.

\end{document}